\documentclass[aps,pra,reprint,superscriptaddress,longbibliography]{revtex4-2}
\usepackage[T1]{fontenc}
\usepackage{lmodern}
\usepackage{amsmath,amssymb,bm,graphicx,booktabs,microtype}
\usepackage{hyperref}
\usepackage{orcidlink}
\hypersetup{colorlinks=true,linkcolor=blue,citecolor=blue,urlcolor=blue}
\newcommand{\tr}{\operatorname{Tr}}
\begin{document}
\title{Coherent Conversion of an Interacting Dark-State-Polariton Condensate into a Photonic Bose–Einstein Condensate}
\author{G. D. de Moraes Neto~\orcidlink{0000-0003-4273-8380}}
\email{gentil.neto@hibiuh.edu.cn}
\affiliation{Department of Fundamental Sciences, Hainan Bielefeld University of Applied Sciences, Danzhou, Hainan 578101, China}
\author{M. A. de Ponte}
\affiliation{Universidade Estadual Paulista (UNESP), Campus Experimental de Itapeva, 18409-010, Itapeva, S\~ao Paulo, Brazil}
\author{N. G. de Almeida}
\affiliation{Instituto de F\'isica, Universidade Federal de Goi\'as, 74001-970, Goi\^ania, Goi\'as, Brazil}
\author{Miled H. Y. Moussa}
\affiliation{Instituto de F\'isica de S\~ao Carlos, Universidade de S\~ao Paulo, S\~ao Carlos, 13560-970, SP, Brazil}

\begin{abstract}
An interacting condensate can retain an observable imprint of its preparation after the interactions are switched off. For light, accessing this regime requires separating the formation of a photon condensate from the material medium that redistributes its energy. We investigate a multimode-cavity scheme in which an interacting dark-state-polariton condensate is coherently converted into a confined photonic Bose–Einstein condensate as the atom–cavity coupling is removed. The polariton mass, matter fraction, and particle number control the condensate occupation during preparation. Remarkably, changing the mass also modifies the condensate's response to repulsion through the spatial overlap of its orbitals, even when the entire single-particle spectrum is held fixed. Ideal mode-preserving conversion conserves the full natural-orbital occupation spectrum, while finite-time few-body calculations show that additional occupation changes during transfer are much smaller than those established during interacting preparation. Once decoupled from matter, the photons evolve under the free cavity Hamiltonian while retaining correlations acquired in the interacting gas. Interference between cavity modes produces oscillations in the photon density, while resonant pair correlations persist after interaction removal and distinguish the resulting photon state from a quadratic thermal state with the same one-body density matrix.
\end{abstract}
\maketitle

\section{Introduction}

In trapped quantum gases, changing the interactions sets an already prepared many-body state into motion. Collapses and revivals of coherence, collective oscillations, and evolving correlations reveal how the state was organized before that change~\cite{Greiner2002,Polkovnikov2011,Schemmer2018,Bougas2019}. For light, preparation and subsequent evolution are more closely tied to the surrounding medium. Photons in ordinary optical cavities interact weakly and continually escape, so maintaining and equilibrating a many-photon population calls for a source of energy redistribution~\cite{Carusotto2013,BlochCarusottoWouters2022}.

Photon Bose--Einstein condensation demonstrates how matter can supply the energy redistribution that light lacks on its own. In dye-filled microcavities, repeated absorption and emission allow photons to approach a Bose distribution and accumulate in the lowest transverse mode~\cite{Klaers2010NP,Klaers2010Nature,Schmitt2015PRA}. Related routes involving laser-cooled atoms and semiconductor cavities broaden the settings in which this balance of confinement, thermalization, and loss can be explored~\cite{Wang2019,Pieczarka2024}. The exchange with matter also shapes the condensate after it forms. Photon-number fluctuations depend on the reservoir~\cite{Klaers2012PRL,Schmitt2014PRL}, while pumping, redistribution, and loss influence its approach to equilibrium~\cite{Kirton2013,Kirton2015}. Temporal coherence can respond strongly to correlations between modes even when the population of the condensed mode changes little~\cite{Schmitt2016PRL,Tang2024}. The medium therefore influences both the formation of the condensate and its subsequent dynamics. Preparing the photon gas through matter and then removing that coupling inside the cavity would separate these roles, allowing the inherited many-body structure to evolve without continued reservoir exchange.

\begin{figure*}[t]
\centering
\includegraphics[width=0.98\textwidth]{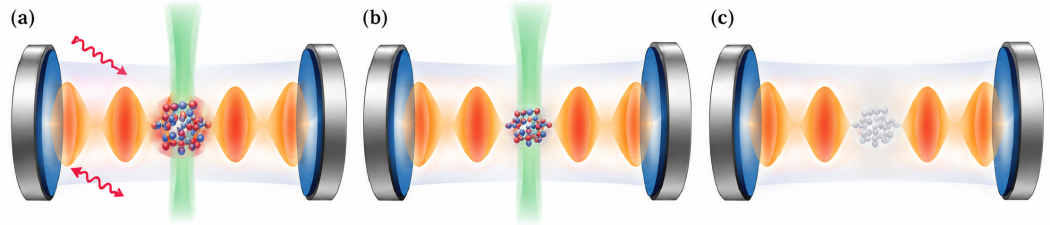}
\caption{Preparation and coherent isolation of a photon condensate in a multimode Fabry--P\'erot-type cavity. Orange lobes represent the longitudinal standing-wave intensity. (a) A polariton gas is prepared at an antinode with optical control (green) and Raman-assisted energy redistribution (red arrows). (b) After reservoir switch-off, the atoms move toward a neighboring node while the control remains active. (c) The residual collective coupling is removed before switching off the control, leaving photons confined and atoms decoupled. The antinode-to-node displacement is one-quarter of the optical wavelength, $\lambda_{\rm opt}/4$; distances are schematic.}
\label{fig:process}
\end{figure*}

Electromagnetically induced transparency (EIT) offers coherent control of the material component itself. A photon and a long-lived collective atomic excitation form a dark-state polariton whose composition follows the applied optical control field~\cite{Fleischhauer2000,Fleischhauer2002,Fleischhauer2005RMP}. In stationary-light geometries, these polaritons acquire a quadratic dispersion with tunable effective mass~\cite{Zimmer2008}. Their atomic component also mediates interactions, opening a route to collision-assisted condensation; in the original stationary-light proposal, the condensate is observed by releasing it as propagating light~\cite{Fleischhauer2008}. Interacting-polariton theories describe the resulting many-body dynamics and optical correlations~\cite{Kiffner2010,KimKim2026}, while experiments have resolved interaction-induced phases and changes in transverse propagation~\cite{Kim2021,KimYu2023}. Spatial control of the effective mass provides an additional way to confine dark polaritons in two dimensions~\cite{Chen2026}. The same ingredients also permit a different final state: confined cavity photons evolving after the atomic coupling has been removed.

In the two-dimensional multimode cavity considered here, the effective mass determines the transverse spectrum and orbital extent, and the matter fraction controls the interaction acquired during preparation. We examine Raman-assisted redistribution as a candidate number-conserving preparation channel and describe the interacting equilibrium target with a fixed-number canonical state. Once the condensate has been prepared, the reservoir is switched off and the light--matter coupling is coherently reduced. The residual atom--cavity coupling is brought to zero while the optical control remains finite; switching off that control then leaves the photons confined and dynamically decoupled from matter (Fig.~\ref{fig:process}).

The photon gas emerging from this sequence retains an interaction-modified condensate. We characterize its occupation through the largest eigenvalue of the one-body density matrix~\cite{PenroseOnsager1956} and examine how that occupation depends on mass, matter admixture, and particle number. Ideal mode-preserving conversion conserves the natural-orbital occupations, while the projected finite-time dynamics introduce only a small additional change in the regimes studied here. After isolation, the photons evolve under the free cavity Hamiltonian. Their evolving coherences and persistent resonant pair correlations then reflect the interacting preparation, independently of ongoing material exchange.

\section{Dark-state polaritons and matter-assisted preparation}

Within a selected longitudinal standing-wave family, the cavity supports a two-dimensional ladder of transverse harmonic modes. Each mode $q=(n_x,n_y)$, with nonnegative oscillator indices $n_x$ and $n_y$, couples to an atomic optical polarization and a long-lived collective spin excitation. We denote their annihilation operators by $a_q$, $P_q$, and $S_q$, respectively. The cavity--polarization coupling $G(t)$ and the control-field Rabi frequency $\Omega_c(t)$ connect the three excitations and are taken to be uniform over the retained modes; in a rotating frame, $\epsilon_{\gamma q}$ and $\epsilon_{sq}$ are the photon and spin-wave frequencies, and $\Delta_q$ is the optical-polarization detuning. The corresponding EIT Hamiltonian is~\cite{Fleischhauer2005RMP}
\begin{align}
H_{\mathrm{EIT}}/\hbar=\sum_q\bigl[&\epsilon_{\gamma q}a_q^\dagger a_q+\Delta_qP_q^\dagger P_q+\epsilon_{sq}S_q^\dagger S_q\notag\\
&+G(t)(P_q^\dagger a_q+\mathrm{H.c.})\notag\\
&+\Omega_c(t)(P_q^\dagger S_q+\mathrm{H.c.})\bigr].\label{eq:eit}
\end{align}
On two-photon resonance, the photon and spin-wave amplitudes can combine so that their couplings to the excited polarization cancel. The resulting dark excitation is $D_q=\cos\theta\,a_q-\sin\theta\,S_q$, with $\tan\theta=G/\Omega_c$. Its photon and matter fractions are $f_\gamma=\cos^2\theta$ and $f_m=\sin^2\theta=1-f_\gamma$. This bosonic description applies when the number of collective spin excitations is small compared with the available atoms.

The photon component supplies the transverse dispersion of the dark polariton. For an effectively dispersionless spin wave and weak dark--bright mixing, the cavity's transverse harmonic motion projects onto the dark branch as
\begin{equation}
m_D=\frac{m_\gamma}{f_\gamma},\qquad \Omega_D=f_\gamma\Omega_\gamma,\label{eq:mass}
\end{equation}
where $m_\gamma$ is the effective photon mass and $\Omega_\gamma$ is its transverse trapping frequency. The dark-polariton shell $j=n_x+n_y$ then lies $\hbar\Omega_Dj$ above the ground mode and contains $j+1$ orbitals. At fixed cavity trapping curvature, changing the optical mixing fraction shifts these spacings while leaving the oscillator length $a_{\mathrm{ho}}=\sqrt{\hbar/(m_D\Omega_D)}$ unchanged. Changing the underlying photon mass also changes that length, and therefore the overlap of transverse orbitals within a finite-range interaction. These two controls consequently affect the prepared gas in different ways.

The ideal-gas occupation of the transverse ladder provides a scale for the finite-number condensation crossover. At particle number $N$, the excited-state saturation condition is
\begin{equation}
N=\sum_{j\geq1}\frac{j+1}{\exp(jx_c^{(N)})-1},\label{eq:cross}
\end{equation}
where $x_c^{(N)}$ is the dimensionless harmonic spacing at this crossover reference. Actual occupations and interaction responses are calculated in the fixed-$N$ canonical ensemble~\cite{Balazs1998,WeissWilkens1997}. Writing $\beta=(k_BT)^{-1}$ and $x=\beta\hbar\Omega_D$, we parameterize the photon mass by $r=m_\gamma/m_*$, where the mass unit $m_*$ is fixed by $\beta\hbar\Omega_\gamma(m_*)=x_*$. This gives $x=x_*f_\gamma/\sqrt r$. For the numerical mass convention, $x_*=x_c^{(N_*)}\simeq0.097$ at $N_*=200$; the corresponding saturation reference at any $N$ is $r_c(N)=[x_*f_\gamma/x_c^{(N)}]^2$. The same mass unit is used when the particle number changes.

Energy redistribution within the dark branch can be supplied by an auxiliary far-detuned Raman transition. After elimination of an intermediate excited level, its coupling to a bath mode $b_\ell$ contains $a_q^\dagger S_{q'}b_\ell^\dagger+\mathrm{H.c.}$; projection onto the dark branch gives
\begin{equation}
 a_q^\dagger S_{q'} \longrightarrow
 -\sqrt{f_\gamma f_m}\,D_q^\dagger D_{q'},
 \qquad [N_D,D_q^\dagger D_{q'}]=0,
 \label{eq:ramanproj}
\end{equation}
with $N_D=\sum_qD_q^\dagger D_q$. The exchange transfers a polariton between transverse orbitals while conserving the total polariton number. With bath rates balanced at the free dark-polariton energy differences, it produces a boson-enhanced redistribution model. In the illustrative free-orbital kinetic closure, reaching a Bose-profile distance below $0.05$ while retaining more than $90\%$ of the mean population requires redistribution faster than loss by a factor of order $6\times10^2$ (Appendix~\ref{sec:SRaman}). That comparison measures free-polariton occupation redistribution and survival, not preparation of an interacting Gibbs state. In the interacting gas, equilibration instead involves transitions at the many-body energy differences of the interacting Hamiltonian. Spatially common Raman coupling can also leave relative-motion sectors disconnected, so equilibration of the interacting gas requires channels that connect those sectors as well as resolve their transition energies (Appendix~\ref{sec:SRaman}). The equilibrium preparation used below assumes that connectivity.

The material component also supplies the collisions that modify the condensate before conversion. We represent them by a repulsive Gaussian potential between dilute spin-wave excitations, with physical range $\sigma$ and microscopic coupling $g_m$. This is a finite-range effective interaction in the two-dimensional harmonic basis~\cite{Doganov2013}, motivated more broadly by matter-mediated optical nonlinearities~\cite{Peyronel2012,Firstenberg2013}. Projecting two incoming and two outgoing spin-wave operators into the dark branch weights the pair interaction by $f_m^2$. In oscillator units, the resulting fixed-$N$ Hamiltonian reads
\begin{align}
\frac{H_D}{\hbar\Omega_D}&=A+\frac{\lambda}{N-1}S,\label{eq:intH}\\
A&=\sum_q j_qD_q^\dagger D_q,\notag\\
S&=\frac12\sum_{abcd}V_{abcd}
D_a^\dagger D_b^\dagger D_cD_d.\notag
\end{align}
Here $j_q=n_x+n_y$, $A$ counts transverse excitation quanta, $S$ is the pair-scattering operator, and $V_{abcd}$ are the Gaussian oscillator matrix elements normalized by $V_{0000}=1$. Their dependence on the orbital width is retained in the dimensionless interaction strength,
\begin{equation}
\lambda=\frac{(N-1)f_m^2g_m J_{0000}}{\hbar\Omega_D},\quad
J_{0000}=\frac{1}{2\pi(a_{\mathrm{ho}}^2+\sigma^2)}.\label{eq:lambda}
\end{equation}
For fixed $g_m$ and $\sigma$, increasing the particle number strengthens the collective interaction parameter proportionally to $N-1$; the coefficient $\lambda/(N-1)$ retains the underlying pairwise coupling. Mass changes modify both level spacing and orbital overlap, while the matter fraction changes the projected interaction as well as the spacing. In the calculations, the reference composition is $f_m=0.05$, and $\lambda_0$ denotes $\lambda$ at $(r,f_m)=(0.4,0.05)$ for the chosen $N$ and the same bare $g_m$. The Gaussian range at this mass setting is $\sigma/a_{\mathrm{ho}}=0.8$.

The equilibrium preparation used in the many-body calculations is the fixed-number interacting Gibbs state
\begin{equation}
\rho_i=Z_N^{-1}\Pi_N e^{-\beta H_D}\Pi_N.\label{eq:rhoinput}
\end{equation}
Here $\Pi_N$ projects onto the $N$-polariton sector and $Z_N=\tr(\Pi_Ne^{-\beta H_D})$ normalizes the state. A number-conserving reservoir that exchanges energy at the interacting transition frequencies and connects the relevant many-body states can relax toward this distribution; an explicit detailed-balance generator for that equilibrium target is given in Appendix~\ref{sec:Sexact}. Reservoir spectra and competing losses determine how closely such an equilibrium state is approached~\cite{Kirton2015}. The condensate occupation and correlations of $\rho_i$ therefore characterize the many-body structure carried into the subsequent coherent conversion.

\begin{figure*}[t]
\centering
\includegraphics[width=0.94\textwidth]{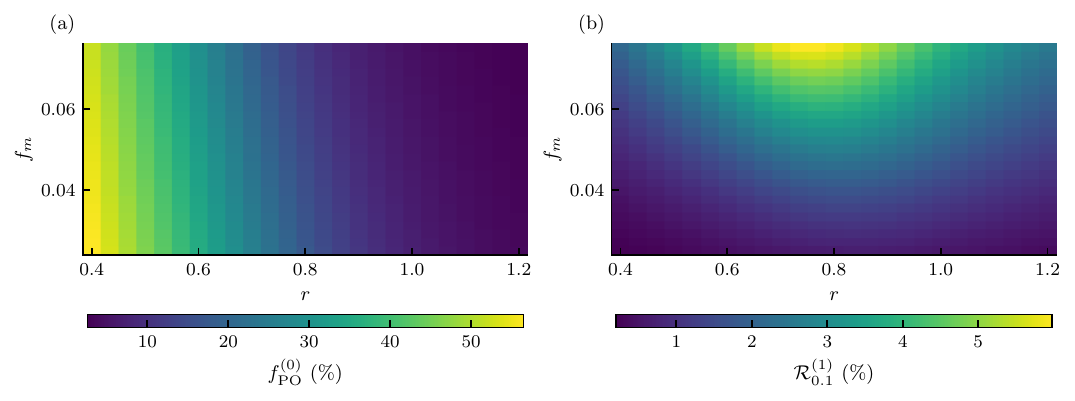}
\caption{Mass and matter content control both the free condensate occupation and its response to repulsion, for the representative $N=200$ fixed-number gas. (a) Free Penrose--Onsager occupation fraction. (b) Relative first-order depletion, $\mathcal R_{0.1}^{(1)}=-100(0.1)D/f_{\mathrm{PO}}^{(0)}$, shown in percent, with $D=\left.\partial f_{\mathrm{PO}}/\partial\lambda_0\right|_0$. The bare Gaussian interaction strength and range remain fixed; $\lambda_0=0.1$ refers to the reference setting $r=0.4$, $f_m=0.05$. The colors in (b) display a zero-coupling susceptibility scaled by that reference value. Canonical thermal orbitals are retained through shell $j\le120$.}
\label{fig:massmap}
\end{figure*}

\begin{figure}[t]
\centering
\includegraphics[width=\columnwidth]{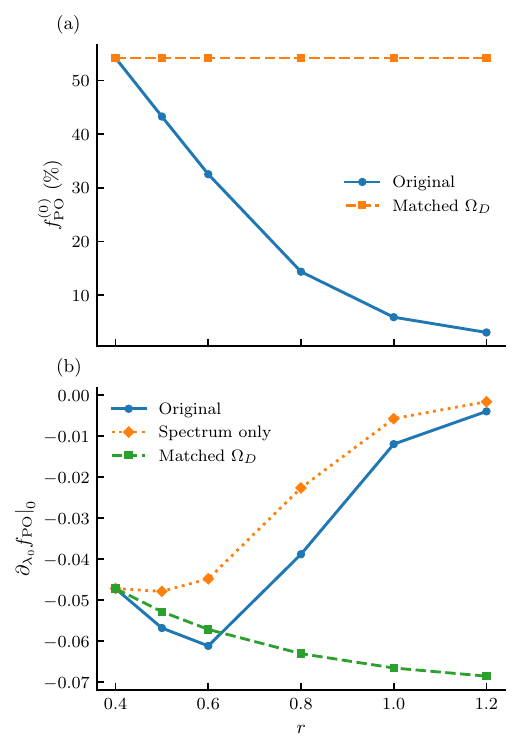}
\caption{Mass-control interpretation, $N=200$ and $f_m=0.05$. (a) Free canonical condensate fraction for the original fixed-curvature mass family and the matched-frequency counterfactual. (b) Bare-coupling susceptibility $D$ for the original family, a spectral-only mathematical control, and the matched-frequency family. The latter holds the entire free harmonic spectrum fixed, not the interaction tensor. All interaction curves are first-order derivatives at zero reference coupling.}
\label{fig:Smatched}
\end{figure}

\section{Mass-dependent condensate occupation and interactions}
Repulsive interactions redistribute particles among transverse orbitals during preparation and can change the shape of the most strongly occupied orbital. The population of the free oscillator ground mode then need not equal the occupation of the interacting condensate. The appropriate quantity is the largest eigenvalue of the one-body density matrix~\cite{PenroseOnsager1956},
\begin{equation}
\gamma_{ab}=\langle D_a^\dagger D_b\rangle,\qquad
f_{\mathrm{PO}}=\frac{\lambda_{\max}(\gamma)}{N}.\label{eq:PO}
\end{equation}
The associated eigenvector gives the condensate orbital. A macroscopic eigenvalue identifies condensation in the many-body limit; for a finite gas, $f_{\mathrm{PO}}$ tracks the leading-orbital occupation continuously through the crossover, including the weakly occupied normal-side region.

At weak interaction, the initial change in this occupation can be obtained directly from the free fixed-number ensemble. Its leading orbital is the nondegenerate oscillator ground state, and the first-order shift of the largest eigenvalue is therefore the change in that orbital's population,
\begin{equation}
\left.\frac{\partial f_{\mathrm{PO}}}{\partial\lambda}\right|_0
=-\frac{x}{N(N-1)}\operatorname{Cov}_0(n_{00},S),\label{eq:POslope}
\end{equation}
where $x=\beta\hbar\Omega_D$, $n_{00}=D_{00}^\dagger D_{00}$, and $S$ is the pair-scattering operator in Eq.~\eqref{eq:intH}. The covariance compares the interaction energy of thermal configurations with different ground-orbital populations. Direct and exchange interactions with the thermally occupied modes determine its magnitude and sign (Appendix~\ref{sec:SPO}).

Mass and matter fraction affect this response through different aspects of the polariton gas. At fixed trapping curvature, increasing the photon mass reduces the transverse level spacing and spreads particles over more orbitals. Changing the atomic fraction alters both that spacing and the interaction inherited by the dark branch. The two effects are visible in Fig.~\ref{fig:massmap}, where we characterize the response at fixed microscopic coupling by $D(r,f_m)=\left.\partial f_{\mathrm{PO}}/\partial\lambda_0\right|_0$. At the reference matter fraction $f_m=0.05$, the free leading-orbital occupation decreases from about $54\%$ to $14\%$ and then $3\%$ as $r$ increases from $0.4$ through $0.8$ to $1.2$. The corresponding relative first-order depletions, scaled to a reference $\lambda_0=0.1$, are roughly $0.9\%$, $2.7\%$, and $1.3\%$. Repulsion thus has its largest relative effect at the intermediate of these sampled masses, even as the free occupation falls with mass. The map compares zero-coupling sensitivities; finite-strength occupation changes also depend on higher-order terms, particularly where the locally rescaled interaction exceeds its reference value.

The free spectrum alone cannot account for the mass dependence of the interacting condensate. To separate its contribution from that of the spatial orbitals, we compare the fixed-curvature family with a mathematical control in which the trap curvature changes with mass so that the transverse harmonic spacing stays fixed. The temperature, matter fraction, microscopic Gaussian coupling, and physical interaction range are held constant. The free canonical distribution is then identical at every mass, while the oscillator wavefunctions narrow as the mass increases. A second, spectrum-only control isolates the change in the free occupation distribution without the accompanying spatial-interaction changes.

Figure~\ref{fig:Smatched} makes the distinction visible. In the matched-spectrum family the free leading-orbital occupation stays near $54\%$, yet the magnitude of its zero-coupling interaction susceptibility increases by about $45\%$ between $r=0.4$ and $r=1.2$. The interactions therefore distinguish states with identical single-particle energies and free thermal occupations: changing the orbital extent alters the finite-range pair-scattering matrix elements. The plotted derivatives quantify weak-coupling sensitivity, not a finite-interaction depletion at each mass. The construction of the matched-frequency Hamiltonian and the decomposition into spectral, coupling, and orbital-shape contributions are given in Appendix~\ref{sec:Smasscontrol}. In the opposite limit of an interaction potential effectively constant over the occupied orbitals, the pair energy becomes a fixed-$N$ offset and the condensate occupation no longer responds to it. The observed mass dependence therefore relies on the finite spatial structure of the interaction.

A finite-strength example shows the same mechanism at the level of an interacting state. In the exact $N=3$ Gibbs calculation within six transverse orbitals, the temperature is chosen separately to give the same free $54\%$ leading-orbital occupation as the condensed-side reference. At $r=0.4$ and $\lambda_0=0.1$, repulsion lowers that occupation by about $1\%$, while its orbital profile changes only weakly. This small-system example complements the large-$N$ response calculation, with the temperature and orbital truncation specified in Appendix~\ref{sec:SPO}.

\begin{figure*}[t]
\centering
\includegraphics[width=0.98\textwidth]{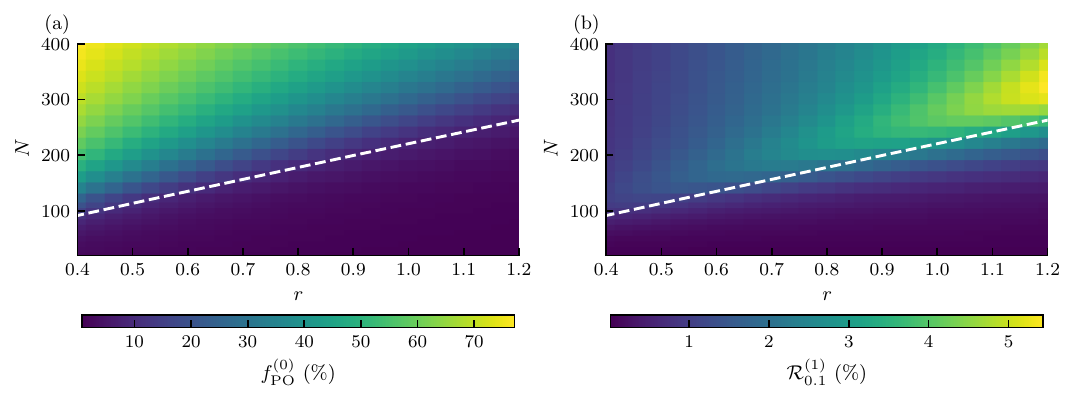}
\caption{Variation of the prepared condensate with excitation number $N$ and mass ratio $r$, at fixed temperature, matter fraction $f_m=0.05$, and microscopic Gaussian coupling and range. (a) Free canonical Penrose--Onsager occupation fraction. (b) Relative first-order depletion, $\mathcal R_{0.1}^{(1)}=-100(0.1)D_N/f_{\mathrm{PO}}^{(0)}$, in percent, where $D_N$ is the bare-coupling susceptibility referenced to $N_*=200$ and $\lambda_{N_*}=0.1$. The dashed curve is the finite-size saturation reference $r_c(N)$; the thermal basis includes shells $j\le120$. Panel (b) compares zero-coupling responses throughout the grid.}
\label{fig:Nscaling}
\end{figure*}

\subsection{Particle-number dependence at fixed microscopic interaction}
\label{sec:Nscaling}

The number of excitations determines how densely the transverse spectrum is populated. At fixed temperature, cavity curvature, and light--matter composition, increasing $N$ moves the finite-size condensation crossover to larger mass. The excited-state saturation reference shifts from $r_c\simeq0.17$ at $N=40$ to about $1.85$ at $N=400$. At the same fixed mass $r=0.4$, the free leading-orbital occupation rises from roughly $3\%$ to $77\%$ across this population range. The smooth boundary in Fig.~\ref{fig:Nscaling}(a) traces this finite-size crossover; representative crossover masses are listed in Table~\ref{tab:Ncrossover}.

Increasing particle number also gives each excitation more interaction partners. With the same microscopic $g_m$, interaction range, geometry, and atomic fraction, the collective dimensionless coupling grows proportionally to $N-1$ [Eq.~\eqref{eq:lambda}]. We denote the coupling at particle number $N$ by $\lambda_N$, and its value at the reference number $N_*=200$ by $\lambda_{N_*}$. At the same bare coupling, the condensate susceptibility is
\begin{equation}
 D_N(r)=\left.\frac{\partial f_{\rm PO}^{(N)}}{\partial\lambda_{N_*}}\right|_0
 =\frac{N-1}{N_*-1}G(r,f_m)
 \left.\frac{\partial f_{\rm PO}^{(N)}}{\partial\lambda_N}\right|_0,
 \label{eq:Nresponse}
\end{equation}
where $G(r,f_m)=\lambda_N(r,f_m)/\lambda_N(0.4,0.05)$ at fixed bare coupling carries the mass--matter dependence of the interaction normalization (Appendix~\ref{sec:SPO}). Figure~\ref{fig:Nscaling}(b) then shows how the relative weak-interaction response changes across the same $(N,r)$ plane. Both the free occupation and the collective interaction scale evolve with particle number, so the regions of strong condensation and strong relative sensitivity need not coincide. As in Fig.~\ref{fig:massmap}, the displayed percentages represent first-order susceptibilities, with their finite-strength range controlled by the local collective coupling.

One can also adjust the temperature at each particle number to obtain the same free leading-orbital occupation. The matched-condensate calculation retains a dependence on $N$ through canonical finite-size statistics and the number of interacting pairs (Appendix~\ref{sec:SNscaling}). Together, the mass, matter, and particle-number controls determine the occupation and orbital structure of the prepared state. These are the many-body properties carried into the coherent conversion considered next.

\begin{figure}[t]
\centering
\includegraphics[width=\columnwidth]{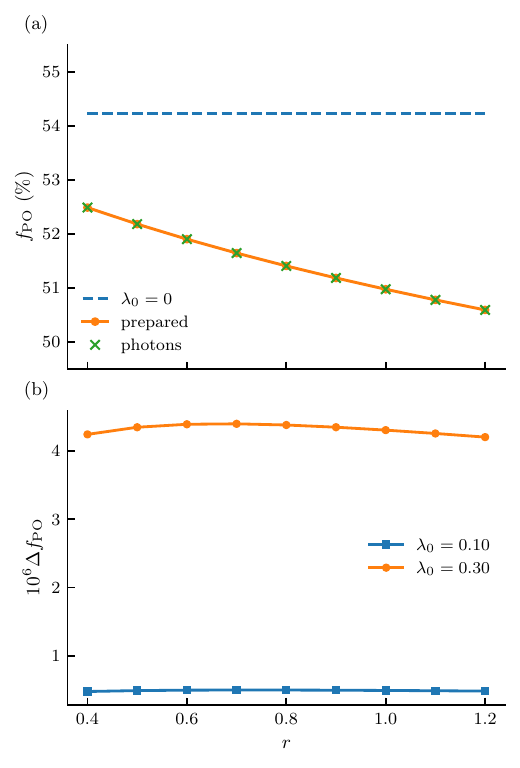}
\caption{Condensate occupation during coherent conversion, calculated for three bosons in the six-orbital projected model. (a) Repulsive preparation lowers the condensate fraction relative to the noninteracting state, while the prepared and photon-output fractions nearly coincide. (b) The much smaller conversion-induced change, resolved on a $10^{-6}$ scale for reference interactions $\lambda_0=0.10$ and $0.30$. At each mass, the small-system temperature is chosen to give the same initial free condensate fraction (about $54\%$); the physical oscillator spacing and local interaction follow the original mass-control and fixed-microscopic-coupling prescriptions. The conversion lasts $1.25\,\mu\mathrm{s}$.}
\label{fig:retention}
\end{figure}

\section{Coherent transfer to an isolated photon condensate}
After the polariton gas has acquired its interacting condensate occupation and correlations, the material component can be withdrawn while the photons remain in the cavity. The preparation reservoir is switched off before conversion begins. Increasing the EIT control then makes the dark excitation progressively more photonic, while translating the atomic ensemble toward a longitudinal standing-wave node suppresses its overlap with the cavity field. For an ensemble of finite axial width, this displacement leaves a residual collective atom--cavity coupling $G_{\rm res}$, inherited from $G(t)$ in Eq.~\eqref{eq:eit}. The final step brings $G_{\rm res}$ to zero \emph{while the control field is still on}; only then is $\Omega_c$ switched off. At that endpoint the dark excitation is the cavity photon, $D_q=a_q$, and the matter-mediated interaction has disappeared. The confined light subsequently evolves with the free transverse photon Hamiltonian, subject to ordinary cavity attenuation.

In the ideal mode-preserving limit, the condensate occupation survives the transfer. More generally, $D_q^\dagger\mapsto\sum_p U_{pq}a_p^\dagger$, where $U$ is a unitary transformation of the transverse orbitals. The one-body density matrix of the output field is $\gamma_\gamma=U^*\gamma_DU^{\mathsf T}$. Its eigenvalues are therefore identical to those of the prepared polariton gas: every natural-orbital occupation, including $Nf_{\mathrm{PO}}$, survives for arbitrary fixed particle number and interaction strength. The corresponding second-quantized unitary also preserves the purity of the full many-body state. During subsequent loss-free evolution under the free photon Hamiltonian, the natural-orbital spectrum remains constant even as coherences between different transverse energies acquire relative phases. The photon field can consequently retain a condensate without being in thermal equilibrium with its final Hamiltonian.

To follow the removal of the interaction over a finite interval, we use the projected dark-manifold Hamiltonian
\begin{align}
H_D(t)&=\hbar\Omega_D(0)\left[h(t)A
+\frac{\lambda w(t)}{N-1}S\right],\label{eq:ramp}\\
h(t)&=\frac{f_\gamma(t)}{f_\gamma(0)},\qquad
w(t)=\left[\frac{f_m(t)}{f_m(0)}\right]^2.\notag
\end{align}
Here $A$ and $S$ are the transverse harmonic and Gaussian-interaction operators introduced in Sec.~II. The factor $h(t)$ follows the increase in the photonic fraction and hence the transverse energy spacing, whereas $w(t)$ tracks the weakening of matter-mediated pair scattering. The active optical conversion lasts $1.25\,\mu\mathrm{s}$ in the prescribed trajectory; the preceding reservoir switch-off adds $0.10\,\mu\mathrm{s}$ to the complete post-preparation sequence (Appendix~\ref{sec:SRaman}).

Two dynamical scales govern this description. The optical dark--bright separation is much larger than the transverse oscillator spacing, allowing the optical composition to follow its controlled dark branch while transverse many-body occupations change more slowly. The quoted bright-state gap characterizes the specified single-particle EIT model. Mode preservation also requires the transverse overlap associated with atomic motion to remain diagonal in the free-mode basis, or to produce negligible off-diagonal mixing. Axial factorization alone does not impose that condition. The corresponding spatial-overlap and spectral benchmarks are collected in Appendix~\ref{sec:SRaman}. For a noninteracting, mode-diagonal reference, the unchanged Bose occupations can be expressed using the rescaled temperature $T_{\rm out}\simeq T_{\rm in}/f_\gamma$ as the harmonic spacing changes. The interacting output instead retains its prepared off-diagonal coherences.

The prepared interaction correction and the additional change during conversion occur at different orders. The free fixed-number Gibbs state is a function of $A$ and commutes with each free-orbital number operator. As a result, the first-order occupation current generated by the time-dependent interaction vanishes. The first-order \emph{equilibrium} change in condensate occupation found in Sec.~III is thus retained through the ramp. When the leading natural orbital is nondegenerate,
\begin{equation}
f_{\mathrm{PO,out}}-f_{\mathrm{PO,in}}=O(\lambda^2).\label{eq:retention}
\end{equation}
The susceptibilities mapped in Figs.~\ref{fig:massmap} and~\ref{fig:Nscaling} therefore describe both the prepared condensate and the photon output to first order within the projected model. Higher-order reorganization can depend on particle number, particularly as the collective coupling grows with the number of interacting partners at fixed microscopic strength.

The difference between preparation and conversion is especially clear in the finite-strength three-boson calculation. At matched initial condensation, repulsion reduces the leading natural-orbital occupation by about $3.2\%$, whereas the subsequent conversion changes that prepared occupation by less than $0.001\%$ over the sampled mass interval (Fig.~\ref{fig:retention}). The exact propagation uses the three-boson Hilbert space within six transverse orbitals. Temperature is adjusted to match the initial free condensate fraction, while the physical oscillator spacing follows the original mass-control protocol. Increasing the ramp duration allows more transverse reorganization (Fig.~\ref{fig:Sduration}; Appendix~\ref{sec:SPOconversion}).

The small conversion correction also persists for $N=2$--5 at matched initial free condensate fractions (Table~\ref{tab:Nconversion}). Within the six-orbital models, it remains below $0.002\%$ at fixed collective interaction strength. Holding the \emph{microscopic} coupling fixed instead weakens the interaction in these few-body examples relative to the larger-population reference, producing still smaller corrections (Appendix~\ref{sec:SNscaling}). Together, the ideal result and finite-time calculations show how the interacting condensate can remain in the photon field after the material coupling is removed. Its inherited off-diagonal and pair correlations then govern additional dynamics during free intracavity storage.

\begin{figure}[t]
\centering
\includegraphics[width=\columnwidth]{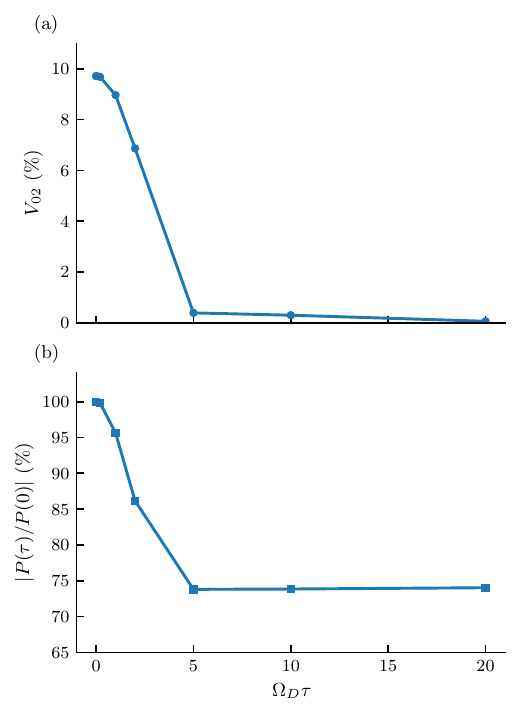}
\caption{Interaction-only smoothstep ramp in the complete six-orbital $N=3$ model at local $\lambda=0.75$. (a) Reduction of one-body interference visibility with ramp duration. (b) Retention of the resonant pair amplitude. The longest interaction-only ramp lasts about $105\,\mu$s at the condensed-side reference spacing, compared with $1.25\,\mu$s for the staged EIT conversion. The curves describe coherent dark-manifold evolution.}
\label{fig:selectivity}
\end{figure}

\section{Correlations inherited by the intracavity photons}
The photon field retains more than the condensate occupation prepared in the interacting polariton gas. Pair scattering between oscillator orbitals $00$, $10$, and $20$ creates two distinct coherences: a one-body amplitude between different energies and a pair amplitude between degenerate two-particle configurations,
\begin{equation}
C_{02}=\langle D_{00}^\dagger D_{20}\rangle,\qquad
P=\langle D_{00}^\dagger D_{20}^\dagger D_{10}^2\rangle.\label{eq:obs}
\end{equation}
In the harmonic spectrum, $E_{20}-E_{00}=2\hbar\Omega_D$ while $E_{00}+E_{20}=2E_{10}$. The first coherence therefore produces a beat at $2\Omega_\gamma$ after isolation; the resonant pair operator commutes with the free harmonic Hamiltonian and retains a constant expectation value.

The normalized one-body interference visibility is $V_{02}=2|C_{02}|/(\langle n_{00}\rangle+\langle n_{20}\rangle)$, where $n_q=D_q^\dagger D_q$ before conversion and the corresponding photon occupations are used afterward. Both amplitudes follow from fixed-number canonical response to the repulsive interaction~\cite{Kubo1957}. At the representative $N=200$, $r=0.4$ setting, $P\simeq-5.1\lambda$ and $V_{02}\simeq0.11|\lambda|$ to first order. The locally normalized weak coupling $|\lambda|=0.1$ gives roughly $1\%$ one-body interference visibility. Comparisons at fixed microscopic coupling additionally require the mass-dependent interaction normalization of Sec.~III.

During conversion, a coherence responds according to the energy difference of the configurations it connects. For an operator $O$ with free harmonic Bohr frequency $\omega_0$ and vanishing free-canonical mean, the projected first-order response is
\begin{align}
\langle O(\tau)\rangle&=\lambda\chi_O e^{i\Phi(\tau)}K_O+O(\lambda^2),\notag\\
K_O&=1-i\omega_0\int_0^\tau w(t)e^{-i\Phi(t)}\,dt,\notag\\
\Phi(t)&=\omega_0\int_0^t h(u)\,du.\label{eq:kernel}
\end{align}
Here $\chi_O=\left.\partial_\lambda\langle O\rangle\right|_{\lambda=0}$ is the equilibrium susceptibility; $h(t)$ and $w(t)$ are the oscillator and interaction envelopes defined in Eq.~\eqref{eq:ramp}. The prescribed EIT ramp retains more than $99.8\%$ of the leading $C_{02}$ amplitude at $r=0.4$. For the resonant pair, $\omega_0=0$ gives $K_P=1$ at first order. This cancellation of the free pair frequency suggests a separation between one-body and pair-coherence responses when the interaction is removed: a coherence between distinct single-particle energies accumulates a relative phase, whereas the resonant pair configurations remain degenerate under free harmonic motion.

The distinction persists at finite interaction strength in the complete six-orbital, three-boson calculation of Fig.~\ref{fig:selectivity}. For an interaction-only smoothstep ramp at local $\lambda=0.75$, increasing the ramp duration from zero to $\Omega_D\tau=20$ reduces the one-body visibility from about $9.7\%$ to $0.06\%$, while roughly $74\%$ of the initial pair amplitude remains. At finite interaction strength, the pair amplitude evolves beyond the leading-order resonance result $K_P=1$. The comparison isolates coherent interaction removal in a projected, loss-free model. Its longest ramp is much slower than the prescribed $1.25\,\mu\mathrm{s}$ EIT conversion, and cannot be interpreted as a realistic loss-inclusive storage protocol (Appendix~\ref{sec:Sexact}).

\begin{figure}[t]
\centering
\includegraphics[width=\columnwidth]{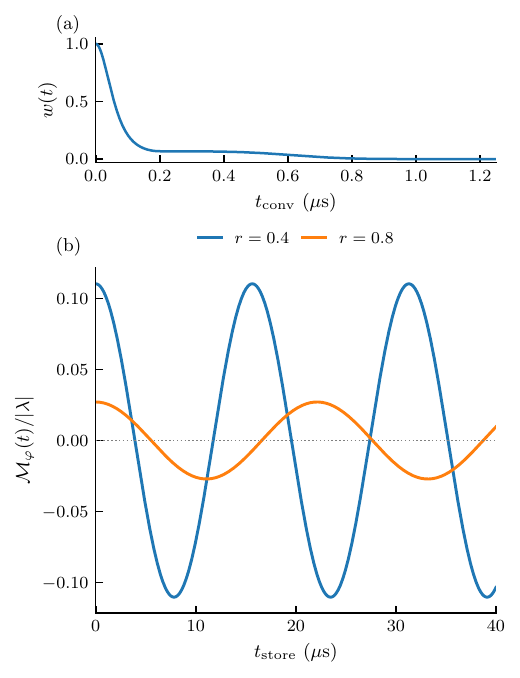}
\caption{Interaction removal and inherited one-body interference. (a) Matter-mediated interaction envelope $w(t)$ during the $1.25\,\mu$s conversion. (b) Phase-referenced contrast $\mathcal M_\varphi(t)/|\lambda|$ during subsequent free-photon storage, measured from a new time origin at the end of conversion. The two mass settings show first-order response per unit \emph{local} coupling $\lambda$ in the loss-free model. A free canonical photon gas gives zero contrast.}
\label{fig:coherence}
\end{figure}

A phase-controlled passive mixer turns the $00$--$20$ coherence into a difference of photon intensities,
\begin{equation}
\mathcal M_\varphi(t)=\frac{\langle n_+(t)\rangle-\langle n_-(t)\rangle}{\langle n_+(t)\rangle+\langle n_-(t)\rangle}
=V_{02}^{\rm out}\cos(2\Omega_\gamma t-\varphi-\phi_0),\label{eq:readout}
\end{equation}
where $a_\pm=(a_{00}\pm e^{i\varphi}a_{20})/\sqrt2$, $\varphi$ is the mixer phase, and $\phi_0=\arg\langle a_{00}^\dagger a_{20}\rangle_{\rm out}$ is the output coherence phase. Time is measured from the start of free-photon storage. Figure~\ref{fig:coherence} shows the interaction envelope during conversion and the subsequent beat for two masses; a free canonical gas gives zero contrast.

The signals measured after isolation also respond differently to attenuation. For a mode-independent linear-loss channel with single-photon survival probability $\eta$, the normally ordered one-body coherence scales as $C_{02}\propto\eta$ and the four-field pair amplitude as $P\propto\eta^2$. Appropriate normalization can remove these uniform attenuation factors, but the probability that all $N$ photons survive scales as $\eta^N$ and can become small even when the mean surviving population remains large. The illustrative loss and spatial-overlap conditions are evaluated in Appendices~\ref{sec:Spreparation} and~\ref{sec:Sspatial}. Mode-dependent loss would require a separate treatment of the output one-body matrix.

The resonant pair amplitude reveals information absent from a quadratic Gibbs state matched to the \emph{complete photon-output one-body matrix}. Reflection symmetry makes that matrix block diagonal between even- and odd-$x$ orbitals. Its matching quadratic generator conserves the number $N_{\rm odd,x}$ in the odd sector, whereas $a_{00}^\dagger a_{20}^\dagger a_{10}^2$ changes that number by two; the pair expectation therefore vanishes in the reference. The interacting pair vertex couples these sectors while preserving overall reflection parity. In the six-orbital $N=3$ output at $r=0.4$, $T=10\,\mu\mathrm K$, and $\lambda=0.30$, the normalized two-particle states differ by trace distance about $0.005$, a maximum unrestricted two-particle measurement contrast of $0.5\%$. Thus two photon states with identical one-body populations and coherences can retain different pair information about their preparation (Appendix~\ref{sec:SPOconversion}).

\section{Discussion and outlook}

During preparation, the atomic component redistributes energy and mediates interactions that shape the polariton condensate. Coherent decoupling then leaves the photons confined, with their subsequent motion governed by the free cavity Hamiltonian. The polariton mass controls the transverse spectrum and the spatial extent of its orbitals, while its atomic fraction determines the interaction strength during preparation. Their combined influence changes with particle number. Even when the free orbital spectrum is held fixed, mass affects the interaction matrix elements and thereby the condensate response.

The transfer carries this interaction-modified structure into the photon field. Ideal mode-preserving conversion conserves the full natural-orbital occupation spectrum, while the projected finite-time dynamics preserve its leading interaction correction. In the exact few-body benchmarks, the additional occupation rearrangement during conversion is much smaller than the change already produced by interacting preparation. The photon condensate retains coherences acquired in the interacting gas: cross-energy coherence generates an optical beat, whereas resonant pair coherence carries information absent from a quadratic Gibbs reference with the same complete one-body density matrix. The medium thus determines properties of the prepared state without remaining coupled during its subsequent intracavity evolution.

A natural extension is to treat the approach to interacting equilibrium and the subsequent conversion within a common many-body dynamics, so that preparation and isolation can be examined together. Finite-strength calculations across increasing particle numbers would connect the present canonical response with the exact few-body retention results. Weak anharmonicity and photon leakage would, in turn, determine the persistence of the inherited coherence during intracavity storage. These extensions would sharpen the connection between preparing a quantum gas through matter-assisted interactions and following its evolution after those interactions have been removed.

\begin{acknowledgments}
The authors acknowledge financial support from CAPES,
CNPq, and FAPESP, Brazilian funding agencies.
\end{acknowledgments}

\section*{Data and code availability}
The data and numerical code supporting the findings of this study are available from the
corresponding author upon reasonable request.

\subsection*{Artificial Intelligence Usage Declaration}
OpenAI ChatGPT (GPT-5.6 Sol) was used
for language editing, bibliographic organization, and code-assisted
analytical and numerical cross-checks. The authors specified the physical
model and validation criteria, checked the resulting calculations against
analytic identities and numerical convergence tests, and assume full
responsibility for the manuscript.
\appendix

\section{Dark-state polaritons, interactions, and mode preservation}
\label{sec:Sprojection}
A cavity photon ($a_q$) and a long-lived collective spin excitation ($S_q$) form a dark polariton by avoiding the short-lived optical polarization ($P_q$). In the EIT Hamiltonian, Eq.~\eqref{eq:eit}, their mixing is governed by the control field and collective cavity coupling~\cite{Fleischhauer2000,Fleischhauer2005RMP}. At two-photon resonance, the normalized dark and bright combinations are
\begin{align}
 D_q&=c a_q-s S_q, & B_q&=s a_q+c S_q,\\
 c&=\Omega_c/\Omega_B, &s&=G/\Omega_B,\quad
 \Omega_B=\sqrt{|G|^2+|\Omega_c|^2}.
\end{align}
For real mode-independent couplings, $a_q=cD_q+sB_q$ and $S_q=-sD_q+cB_q$. The dispersive quadratic terms transform into
\begin{align}
\epsilon_{Dq}D_q^\dagger D_q+\epsilon_{Bq}B_q^\dagger B_q
+\epsilon_{DB,q}(D_q^\dagger B_q+\mathrm{H.c.}),\nonumber\\
\epsilon_{Dq}=c^2\epsilon_{\gamma q}+s^2\epsilon_{sq},\qquad
\epsilon_{DB,q}=cs(\epsilon_{\gamma q}-\epsilon_{sq}).
\end{align}
The last term mixes the dispersive dark and bright sectors. Neglecting it requires the dispersion mismatch to be small compared with the minimum optical bright-state separation, across the thermally occupied transverse manifold. If the spin-wave branch is approximately flat, the projected transverse Hamiltonian is
\begin{equation}
\begin{aligned}
h_D&=f_\gamma\left(\frac{p_\perp^2}{2m_\gamma}
  +\frac12K_\gamma r_\perp^2\right),\\
m_D&=\frac{m_\gamma}{f_\gamma},\qquad
\Omega_D=f_\gamma\Omega_\gamma.
\end{aligned}\label{eq:Sdarkmass}
\end{equation}
The oscillator length $a_{\rm ho}=\sqrt{\hbar/(m_D\Omega_D)}
=\sqrt{\hbar/(m_\gamma\Omega_\gamma)}$ is independent of the EIT mixing angle at fixed photon mass and trap curvature. Thus a homogeneous change of EIT composition preserves the transverse oscillator eigenfunctions within this ideal projection, even though their level spacing changes. This is distinct from changing the photon mass $m_\gamma$, which changes $a_{\rm ho}$.

The interacting extension assumes an additional spin-wave two-body Hamiltonian
\begin{align}
H_{SS}={}&\frac12\int d^2r\,d^2r'\;
 U_\sigma(\bm r-\bm r')\notag\\
&\times S^\dagger(\bm r)S^\dagger(\bm r')
 S(\bm r')S(\bm r),\\
U_\sigma(\bm R)={}&\frac{g_m}{2\pi\sigma^2}
\exp[-|\bm R|^2/(2\sigma^2)],\quad g_m>0.
\label{eq:Sspin}
\end{align}
Here $S(\bm r)$ represents bosonic spin-wave excitations at low density, and $g_m$ has units of energy times area. The finite-range repulsion is an effective interaction supplied by the atomic medium, independent of the quadratic EIT mixing Hamiltonian. Inserting $S_q\simeq-\sqrt{f_m}D_q$ gives $H_{\rm int}\propto f_m^2$. The resulting tensor in a fixed 2D oscillator basis is
\begin{align}
J_{abcd}(\sigma)={}&\iint d^2r\,d^2r'\;
\phi_a^*(\bm r)\phi_d(\bm r)\notag\\
&\quad\times\frac{e^{-|\bm r-\bm r'|^2/(2\sigma^2)}}{2\pi\sigma^2}
\phi_b^*(\bm r')\phi_c(\bm r'),\\
V_{abcd}={}&\frac{J_{abcd}}{J_{0000}},\qquad
\lambda=\frac{(N-1)f_m^2g_mJ_{0000}}{\hbar\Omega_D}.
\label{eq:Stensor}
\end{align}
where $0$ denotes orbital $(0,0)$. Within the dark manifold this tensor gives the quartic Hamiltonian of Eq.~\eqref{eq:intH}. The finite-range Gaussian form is standard in two-dimensional trapped-boson models~\cite{Doganov2013}. The bosonic-spin-wave and dark-branch projections apply when excitation density and dark--bright interaction mixing remain small.

For the normalized 2D harmonic ground state $|\phi_0(\bm r)|^2=(\pi a_{\rm ho}^2)^{-1}e^{-r^2/a_{\rm ho}^2}$, Gaussian convolution gives
\begin{equation}
 J_{0000}=\frac{1}{2\pi(a_{\rm ho}^2+\sigma^2)}.
 \label{eq:Sgroundintegral}
\end{equation}
This normalization ensures that the interaction parameter changes when the transverse mass is changed at fixed physical $g_m$ and $\sigma$.

The many-mode map additionally assumes that the transverse collective coupling remains diagonal in, or has a common eigenbasis with, the free oscillator Hamiltonian. If the atomic density is separable under axial motion, its overlap matrix factorizes as $\mathcal M_{qq'}(t)=A_z(t)\mathcal M^{\perp}_{qq'}$. The matrices at different times therefore commute. Preserving photon energy modes also requires $[\mathcal M^\perp,h_\gamma]=0$, or sufficiently weak off-diagonal coupling. An arbitrary initial fixed-$N$ density operator then maps, conditional on no loss and no bright-state leakage, as
\begin{equation}
 \rho_D\longmapsto U_\phi\rho_DU_\phi^\dagger,
 \qquad D_q^\dagger\longmapsto e^{i\phi_q}a_q^\dagger.
 \label{eq:Smap}
\end{equation}
The diagonal photon occupations and the eigenvalues of the one-body density matrix are conserved by this ideal map; off-diagonal coherences acquire calculable mode phases.

\subsection{Mass control at fixed microscopic interaction}
\label{sec:Snormalization}
To compare different masses at a common physical interaction, the Gaussian range $\sigma$ and bare coupling $g_m$ are held fixed. The mass ratio $r=m_\gamma/m_{\gamma,c}^{(N_*)}$ is defined using the representative population $N_*=200$ and its ideal-gas crossover value $x_c^{(N_*)}\simeq0.097$. With $f_\gamma=1-f_m$,
\begin{equation}
\begin{aligned}
x(r,f_m)&=\beta\hbar\Omega_D
=x_c^{(200)}\frac{1-f_m}{\sqrt r},\\
\frac{\sigma}{a_{\rm ho}(r)}&=0.8\left(\frac{r}{0.4}\right)^{1/4}.
\end{aligned}\label{eq:Sxr}
\end{equation}
for the chosen reference physical range. We fix $g_m$ and $\sigma$ in physical units and take $(r_0,f_{m0})=(0.4,0.05)$.

Because the DSP oscillator spacing contains the photonic fraction, its ideal-gas crossover occurs at a different mass from the bare-photon reference $r=1$. The saturation condition is $x(r,f_m)=x_c^{(200)}$, so
\begin{equation}
 r_c^{\rm DSP}=(1-f_m)^2=f_\gamma^2=0.9025\quad (f_m=0.05).
 \label{eq:Scrossover}
\end{equation}
This location defines the ideal-gas finite-$N$ crossover reference used throughout the canonical calculations.

From Eqs.~\eqref{eq:Sdarkmass} and \eqref{eq:Sgroundintegral},
\begin{equation}
\frac{\lambda(r,f_m)}{\lambda_0}
 =\left(\frac{f_m}{0.05}\right)^2
 \frac{1+0.8^2}{\sqrt{0.4/r}+0.8^2}
 \sqrt{\frac{r}{0.4}}\,\frac{0.95}{1-f_m}
 \label{eq:Slambda}
\end{equation}
with $\lambda_0=\lambda(0.4,0.05)$. The first fraction is the spin-wave projection, the second is the change in the finite-range ground-orbital integral, and the last two account for the transverse energy spacing. This normalization applies to comparisons at fixed underlying bare interaction; it is \emph{not} implicit in the derivative $\partial_\lambda$ plotted separately at each $r$.

At each mass, the correlation susceptibilities $v_1(r)$ and $\partial_\lambda P|_0$ are evaluated at the local oscillator spacing and relative interaction range. For a chosen reference strength $\lambda_0$, the leading-order amplitudes are $V_{02}\simeq v_1(r)|\lambda(r,f_m)|$ and $P\simeq \lambda(r,f_m)\partial_\lambda P|_0$. These expressions describe the weak-coupling slope. The reference value $\lambda_0=0.1$ specifies the illustrative interaction scale used in the plots.

\section{Fixed-number thermal response and correlation functions}
\label{sec:Slinear}
To determine how a weak interaction changes equilibrium occupations and coherences at fixed particle number, write the projected Hamiltonian as $H_D=\hbar\Omega_D(A+\lambda W)$, where $A=\sum_qj_qn_q$ counts harmonic excitation energy and $W$ is the dimensionless quartic term of Eq.~\eqref{eq:intH}. The fixed-$N$ thermal reference is $\rho_0=\Pi_Ne^{-xA}\Pi_N/Z_N$, $x=\beta\hbar\Omega_D$. Differentiating the exponential before taking a trace gives the exact imaginary-time response
\begin{equation}
\begin{aligned}
\chi_O&\equiv\left.\frac{\partial\langle O\rangle}
 {\partial\lambda}\right|_0\\
&=-x\int_0^1du\,\Big[
\langle e^{uxA}W e^{-uxA}O\rangle_0\\
&\hspace{4em}-\langle W\rangle_0\langle O\rangle_0\Big].
\end{aligned}\label{eq:Skubo}
\end{equation}
The imaginary-time response~\cite{Kubo1957} retains the correlations among occupations imposed by fixed particle number. It also treats transitions within and between free-energy manifolds. In a many-body occupation basis, if $A|n\rangle=a_n|n\rangle$, an off-diagonal Gibbs response has matrix elements
\begin{equation}
 (\rho'_0)_{nm}=
 \frac{p_n-p_m}{a_n-a_m}\,W_{nm}\quad(a_n\ne a_m),
 \qquad p_n=\frac{e^{-xa_n}}{Z_N},
 \label{eq:Srhoslope}
\end{equation}
with the continuous degenerate limit $(\rho'_0)_{nm}=-xp_nW_{nm}$ for $a_n=a_m$, $n\ne m$. For diagonal entries the normalization correction is $(\rho'_0)_{nn}=-xp_n(W_{nn}-\langle W\rangle_0)$. Here $n,m$ label many-body occupation configurations, not single-particle orbitals.

The canonical partition coefficients follow the fixed-number generating-function and recurrence methods for harmonically trapped ideal bosons~\cite{Balazs1998,WeissWilkens1997}:
\begin{equation}
 \mathcal Z(z;\bm t)=\prod_{q\in\mathcal M}
 \frac{1}{1-z e^{-xj_q+t_q}},\qquad
 Z_N(\bm t)=[z^N]\mathcal Z(z;\bm t),
 \label{eq:Sgenerating}
\end{equation}
where $\mathcal M$ is the complete retained orbital set, including all degeneracies. At $\bm t=0$, ordinary moments follow by differentiating $Z_N(\bm t)$ with respect to the source fields. More directly the excited-state recurrence used in the noninteracting reference is
\begin{align}
 Z_M^{\rm ex}&=\frac1M\sum_{k=1}^M z_1^{\rm ex}(kx)Z_{M-k}^{\rm ex},
 \quad Z_0^{\rm ex}=1,\\
 z_1^{\rm ex}(kx)&=\sum_{j=1}^{j_{\max}}(j+1)e^{-kjx},\quad
 Z_N=\sum_{M=0}^{N}Z_M^{\rm ex}.
 \label{eq:Srec}
\end{align}
In the untruncated ideal harmonic trap $z_1^{\rm ex}(kx)=(1-e^{-kx})^{-2}-1$. The exact ground-mode distribution is $\Pr(N_0=n)=Z_{N-n}^{\rm ex}/Z_N$. For any specified orbital $a$, for example,
\begin{align}
\langle n_a\rangle_0
&=\frac1{Z_N}\sum_{k=1}^{N}e^{-kxj_a}Z_{N-k},\\
\langle n_an_b\rangle_0
&=\frac1{Z_N}\sum_{\substack{k,l\ge1\\k+l\le N}}
 e^{-x(kj_a+lj_b)}Z_{N-k-l}\notag\\
&\hspace{5em}(a\ne b),\\
\frac12\langle n_a(n_a-1)\rangle_0
&=\frac1{Z_N}\sum_{k=2}^N
(k-1)e^{-kxj_a}Z_{N-k}.
\label{eq:Smoments}
\end{align}
The pair response uses the joint occupation distribution of modes $(00,10,20)$, obtained from their occupation weights and the canonical partition function of the remaining orbitals.

The specific pair operator is $O_P=D_{00}^\dagger D_{20}^\dagger D_{10}D_{10}$. Its creation and annihilation configurations satisfy $0+2=1+1$, so $[A,O_P]=0$. Equation~\eqref{eq:Skubo} therefore reduces to
\begin{equation}
\chi_P=-\frac{x V_{0,2,1,1}}{N-1}
\big\langle(n_{00}+1)(n_{20}+1)n_{10}(n_{10}-1)\big\rangle_0.
\label{eq:Spair}
\end{equation}
We use orbital abbreviations $0=(0,0)$, $1=(1,0)$, $2=(2,0)$ in this equation only. The corresponding cross-shell response $\chi_C$ with $O_C=D_{00}^\dagger D_{20}$ has a nonzero free energy difference $2\hbar\Omega_D$. In the real oscillator basis, define
\begin{equation}
 Q_s=V_{0,s,s,2}+V_{0,s,2,s},
 \qquad
 \chi_C=\frac{1}{2(N-1)}\sum_{s\in\mathcal M}Q_s T_s,
 \label{eq:Scross}
\end{equation}
where
\begin{equation}
T_s=\begin{cases}
\langle n_0n_2\rangle_0-\frac12\langle n_0(n_0-1)\rangle_0,&s=0,\\
\frac12\langle n_2(n_2-1)\rangle_0-\langle n_0n_2\rangle_0,&s=2,\\
\langle n_2n_s\rangle_0-\langle n_0n_s\rangle_0,&s\notin\{0,2\}.
\end{cases}
\label{eq:ST}
\end{equation}
Direct and exchange scattering contribute through the two tensor contractions. Their weights are exact fixed-number occupation moments in the ideal reference ensemble. The small-$\lambda$ visibility is
\begin{equation}
 V_{02}=v_1|\lambda|+O(\lambda^2),\qquad
 v_1=\frac{2|\chi_C|}{\langle n_0\rangle_0+\langle n_2\rangle_0}.
 \label{eq:Svis}
\end{equation}
The absolute value means that $v_1$ is the one-sided linear coefficient; $V_{02}$ is not differentiable in the ordinary two-sided sense at $\lambda=0$.

The pair coherence also isolates correlations beyond contractions of the one-body density matrix. In a fixed-number state invariant under $x\mapsto-x$, the orbitals $0,2$ are even and orbital $1$ is odd. Hence $\langle D_0^\dagger D_1\rangle=\langle D_2^\dagger D_1\rangle=0$. Normal, Wick-factorized contributions to $\langle D_0^\dagger D_2^\dagger D_1D_1\rangle$ are $2\langle D_0^\dagger D_1\rangle\langle D_2^\dagger D_1\rangle=0$, and anomalous pair means vanish by number conservation. The nonzero $P$ is therefore a connected four-field coherence relative to this Wick decomposition. The connected amplitude characterizes four-field coherence; its quantum-correlation interpretation requires an additional witness. For example, take normalized amplitudes $u,v,w$ and the fixed-$N$ product condensates $|\phi_\pm;N\rangle\propto(uD_0^\dagger\pm vD_1^\dagger+wD_2^\dagger)^N|0\rangle$. The equal mixture of these two particle-separable states is invariant under orbital-$1$ parity. It has $\langle D_0^\dagger D_1\rangle=\langle D_2^\dagger D_1\rangle=0$ but $P=N(N-1)u^*w^*v^2\ne0$ for nonzero amplitudes. The example shows that a Wick-connected $P$ can also arise in a mixture of particle-separable condensates. Here $P$ is used as a phase-sensitive record of the preparation.

\subsection{Coherence during interaction removal}
\label{sec:Skernel}
During projected conversion, the Hamiltonian and the oscillator and interaction envelopes are those of Eq.~\eqref{eq:ramp}. Let $[H_0(0),O]=\hbar\omega_0 O$ for a single free Bohr-frequency component $O$ satisfying $\langle O\rangle_0=0$. Expanding the initial Gibbs state through Eq.~\eqref{eq:Skubo} and the time-ordered unitary to first order gives a commutator integral. The Kubo--Martin--Schwinger relation for the free canonical state reduces that integral to the response kernel in Eq.~\eqref{eq:kernel}.
The sign convention for $\omega_0$ changes the phase of the kernel but leaves the retained magnitude $|K_O|$ invariant. At fixed $h=1$ and $w(s)=1-3s^2+2s^3$, integration by parts gives the simple interaction-removal filter
\begin{equation}
 K_O=\int_0^1 6s(1-s)e^{-i\omega_0\tau s}ds.
 \label{eq:Ssmooth}
\end{equation}
The cross-shell operator has $|\omega_0|=2\Omega_D(0)$ and hence a duration-dependent amplitude. The resonant pair operator has $\omega_0=0$, so $K_P=1$ at \emph{first order}. Finite interaction strength, anharmonic dispersion and mode-dependent conversion phases produce additional pair-amplitude evolution.

For the prescribed three-stage EIT trajectory, the collective coupling initially satisfies $G/(2\pi)\simeq63$ MHz. The control frequency $\Omega_c/(2\pi)$ rises from 274 to 548 MHz over $0.20~\mu$s, the atomic ensemble moves toward a coupling node over $1.00~\mu$s, and residual $G$ is suppressed over the final $0.05~\mu$s while the control field remains on. The first control-field stage reduces the nonlinear envelope to about $7\%$ of its initial value. The projected mass scales as in Eq.~\eqref{eq:Sdarkmass}, so the same prescribed optical trajectory determines both the nonlinear envelope and the free oscillator phase. The cross-energy coherence retains more than $99.8\%$ of its initial first-order amplitude at $r=0.4$, while the resonant pair amplitude is retained exactly at this order. These amplitudes refer to the coherent dark-manifold evolution; photon attenuation acts on them separately.

During conversion, rotation of the instantaneous dark--bright basis produces a nonadiabatic coupling of order $|\dot\theta|$ and the static dispersion-induced coupling as $|cs(\epsilon_{\gamma q}-\epsilon_{sq})|$. Their combined magnitude must remain small relative to the minimum relevant bright-state eigenvalue separation. The three-component EIT propagation gives a minimum bright-state gap of about 279 MHz for the specified single-particle trajectory. Its extension to finite-density bright-state dynamics requires additional interaction terms. A mode-dependent nonlinear phase during the conversion would have to be included in a beyond-leading-order model.

\section{Finite-interaction dynamics and conditional preparation}
\label{sec:Sexact}
The finite-strength comparison uses all 2D oscillator orbitals with $n_x+n_y\le2$ (six orbitals), the full symmetric Gaussian tensor of Eq.~\eqref{eq:Stensor}, and fixed boson numbers $N=2$--$5$. The Fock-space dimensions are $\binom{N+5}{N}$, namely 21, 56, 126, and 252. Within this finite orbital space, the exact calculation retains the full symmetric interaction tensor and every fixed-number Fock configuration. The canonical state is computed by exact matrix diagonalization of the resulting many-body Hamiltonian. For the $N=3$ example, the inverse thermal oscillator spacing is adjusted to reproduce the \emph{noninteracting} ground-mode population fraction of the 200-boson, $r=0.4$ reference (approximately $0.54$); it is not the physical temperature of the 200-particle reference.

For a smoothstep interaction-only ramp, $w(t)=1-3(t/\tau)^2+2(t/\tau)^3$, propagation solves $\dot\rho=-(i/\hbar)[H_0+\lambda w(t)W,\rho]$ and evaluates the two observables at the end. The different one-body and pair responses are discussed with Fig.~\ref{fig:selectivity} in Sec.~V. The finite-strength propagation is exact \emph{within the six-orbital finite-range Hamiltonian}; its quantitative scope is the six-orbital few-boson truncation. 

The thermal-orbital cutoff has little effect on the condensed-side first-order response at the representative population $N=200$: increasing the thermal cutoff from $j\le80$ to $j\le120$ changes both coherence coefficients by about 0.005\%. At finite interaction strength, spectator scattering produces larger changes. In the complete six-orbital $N=4$ Hamiltonian at $\lambda=0.75$, omitting spectator interactions changes the pair amplitude by about 23\%, while including their conditional mean density reduces the error to about 5\%. The selected spectator-process approximation at larger $N$ captures only its retained scattering channels.

\subsection{Interacting-state preparation and photonic observables}
\label{sec:Spreparation}
Let $|n\rangle$ denote interacting fixed-$N$ eigenstates with energies $E_n$. An abstract number-conserving detailed-balance generator can be written as
\begin{equation}
\begin{aligned}
\dot\rho&=-\frac{i}{\hbar}[H_D,\rho]
+\sum_{m\ne n}\gamma_{m\leftarrow n}
\mathcal D[|m\rangle\langle n|]\rho,\\
\mathcal D[L]\rho&=L\rho L^\dagger
-\tfrac12\{L^\dagger L,\rho\}.
\end{aligned}\label{eq:SL}
\end{equation}
with $\gamma_{m\leftarrow n}/\gamma_{n\leftarrow m}=\exp[-\beta(E_m-E_n)]$ for allowed paired transitions. Detailed balance makes the interacting Gibbs state stationary. A unique attractive Gibbs state further requires that the transition graph connecting the population sectors be irreducible (and that no unaddressed decoherence-free quantum sector persists); uniqueness follows only when the allowed transitions connect the relevant sectors. This generator defines interacting equilibrium as a possible preparation target. Spatially common Raman driving can preserve relative-motion sectors, so its accessible transitions may differ from those of the energy-resolved generator.

The isolated photon output can be analyzed through mode-resolved optical coherence measurements~\cite{Glauber1963}. Coherent interference between the $00$ and $20$ output modes measures the amplitude $\langle a_{00}^\dagger a_{20}\rangle$, with a free-storage beat at $2\Omega_\gamma$. A balanced mode mixer gives the normalized difference signal defined by Eq.~\eqref{eq:readout}; for the free thermal reference its expectation vanishes. The normally ordered four-field amplitude $\langle a_{00}^\dagger a_{20}^\dagger a_{10}a_{10}\rangle$ requires phase-sensitive, mode-resolved fourth-order interference or an equivalent four-wave-mixing correlation measurement; ordinary intensities do not determine its complex phase. Pair coherence by itself characterizes an interacting preparation history; entanglement and phase-space negativity are separate observables.

The present conversion calculations describe conditional loss-free evolution. If each of $N$ particles survives independently with single-particle survival probability $\eta$, the probability of retaining \emph{all} particles is $\eta^N$; at $\eta=0.9$ and $N=200$ it is $\simeq7.1\times10^{-10}$. Mean retained population, condensate fraction and full-state survival probability describe different aspects of the output. The Raman rate model and coherent-control sequence therefore describe distinct stages of the theoretical preparation.

For a mode-independent single-photon loss model with rate $\kappa=2\pi\times1\,\mathrm{kHz}$, the single-particle mean survival and the probability to retain all $N$ excitations are
\begin{equation}
 \eta(\tau)=e^{-\kappa\tau},\qquad p_{N,\mathrm{no\ loss}}=\eta^N.
 \label{eq:Sloss}
\end{equation}
At $\tau=1.25\,\mu\mathrm{s}$, $\eta\simeq0.992$ and $\eta^{200}\simeq0.21$; at $\tau=20/\Omega_D(0)\simeq105\,\mu\mathrm{s}$ for $r=0.4$, $\eta\simeq0.52$ and $\eta^{200}\simeq6\times10^{-58}$. In a linear uniform attenuation channel the normally ordered coherence $C_{02}$ scales as $\eta$, while the four-field amplitude $P$ scales as $\eta^2$; the factors can cancel in appropriately normalized observables. These numbers apply to the specified uniform-loss channel and its normally ordered observables. In particular, the dark-polariton composition and the loss channels need not be constant during the actual EIT ramp. The longest loss-free smoothstep used in Fig.~\ref{fig:selectivity} establishes the selective coherence response of the projected loss-free model.

\section{Interacting-condensate response and mass control}
\label{sec:SPO}
For an equilibrium fixed-$N$ state let $\gamma_{ab}=\langle D_a^\dagger D_b\rangle$ and $N_{\mathrm{PO}}=\lambda_{\max}(\gamma)$. At $\lambda=0$ the free oscillator modes diagonalize $\gamma$, and the ground-mode occupation is the nondegenerate largest eigenvalue in the parameter window considered. Although the interacting natural orbital changes at first order through $\gamma_{ab}$ with $a\ne b$, the first-order shift in the nondegenerate eigenvalue is the shift of the diagonal element $\langle n_{00}\rangle$. For the dimensionless $H/(\hbar\Omega_D)=A+\lambda S/(N-1)$, and $x=\beta\hbar\Omega_D$, the canonical Kubo identity gives
\begin{equation}
 \left.\frac{\partial\langle n_{00}\rangle}{\partial\lambda}\right|_0
 =-\frac{x}{N-1}\bigl[\langle n_{00}S\rangle_0
 -\langle n_{00}\rangle_0\langle S\rangle_0\bigr].
 \label{eq:SPOcov}
\end{equation}
Although $S$ contains off-diagonal scattering, $n_{00}$ commutes with the free Hamiltonian, so only the free-energy-diagonal part of $S$ contributes to this covariance. Direct and exchange terms couple the ground occupation to all thermally populated orbitals; no selected-mode factorization is made. Dividing Eq.~\eqref{eq:SPOcov} by $N$ yields Eq.~\eqref{eq:POslope}. Equivalently the response may be evaluated as a derivative of the free-canonical interaction energy with respect to a fictitious ground-orbital level shift, a method used in the canonical-response calculation.

For the physical control surface, let $(r_0,f_{m0})=(0.4,0.05)$, $s_0=0.8$, and $\lambda_0$ the dimensionless coupling evaluated there at a fixed bare interaction $g_m$. With $G(r,f_m)=\lambda(r,f_m)/\lambda_0$, the map is
\begin{equation}
 D(r,f_m)=\left.\frac{\partial f_{\mathrm{PO}}}{\partial\lambda_0}\right|_0
 =G(r,f_m)\left.\frac{\partial f_{\mathrm{PO}}}{\partial\lambda}\right|_0.
 \label{eq:SphysicalD}
\end{equation}
The canonical calculation retains complete thermal shells through $j\le120$, comprising 7,381 oscillator modes. At the selected high-mass points, increasing the cutoff from 120 to 160 changes the susceptibility by at most about $0.21\%$. The $N=3$ full-Gibbs checks use all six orbitals with $j\le2$ and the original 56-dimensional fixed-number Hilbert space. At the separately matched small-system temperature, the repulsion at $r=0.4$, $\lambda_0=0.1$ reduces the condensate occupation by about $1.1\%$ relative to the free value. The physical $N=3$ and $N=200$ temperature windows remain different.

\subsection{Matched-spectrum comparison}
\label{sec:Smasscontrol}
The harmonic mass family used throughout the primary calculation keeps transverse cavity curvature fixed. With $r_0=0.4$ and $f_{m0}=0.05$, the dimensionless thermal spacing and range are
\begin{align}
 x(r,f_m)&=x_c^{(200)}\,(1-f_m)/\sqrt r,\\
 s_{\mathrm{orig}}(r)&=s_0(r/r_0)^{1/4},\qquad s_0=0.8.
\end{align}
At fixed bare interaction $g_m$ and physical Gaussian range $\sigma$, its dimensionless coupling factor is
\begin{equation}
 G_{\mathrm{orig}}(r,f_m)=\left(\frac{f_m}{0.05}\right)^2
 \frac{1+s_0^2}{\sqrt{r_0/r}+s_0^2}\sqrt{\frac{r}{r_0}}\frac{0.95}{1-f_m}.
 \label{eq:SGorig}
\end{equation}
Here $G_{\mathrm{orig}}(r_0,f_{m0})=1$. In a mathematically controlled matched-frequency comparison, the cavity trap curvature is adjusted to keep $\Omega_D$ fixed as the photon mass is varied, with temperature, bare $g_m$, $f_m$ and $\sigma$ unchanged. Then
\begin{equation}
\begin{aligned}
x_{\rm match}&=x(r_0,f_m),\\
s_{\rm match}&=s_0\sqrt{r/r_0},\\
G_{\rm match}&=G_{\rm orig}(r_0,f_m)
\frac{1+s_0^2}{r_0/r+s_0^2}.
\end{aligned}\label{eq:SGmatch}
\end{equation}
The ideal condensate occupation is therefore independent of $r$ in the matched-frequency family, while the bare-coupling interaction response changes because oscillator functions and the normalized Gaussian interaction matrix change.

For the decomposition underlying Fig.~\ref{fig:Smatched}, the susceptibility change across $0.4\le r\le1.2$ can be attributed to the three dimensionless controls $(x,G,s)$ by averaging over their six possible orders of variation. Thermal spacing contributes $+0.074$, coupling normalization $-0.032$, and the finite-range orbital shape $+0.002$, summing to $\Delta D\simeq+0.043$. This symmetrized attribution compares the effects of changing the three parameters without privileging an order of variation. In the long-range limit $\sigma\gg a_{\rm ho}$, the normalized two-body interaction tends to a constant in the fixed-$N$ sector and its effect on the condensate occupation vanishes. The Gaussian interaction has fixed, finite physical range throughout these comparisons.

\section{Condensate retention and inherited correlations}
\label{sec:SPOconversion}
An ideal single-particle mode-preserving map $D_i^\dagger\rightarrow\sum_j U_{ji}a_j^\dagger$ is implemented on the fixed-$N$ Fock space by the corresponding second-quantized unitary. Its one-body density matrix is unitarily equivalent to that of the prepared polaritons, and hence $\mathrm{spec}(\gamma_{\rm ph})=\mathrm{spec}(\gamma_D)$. The same unitary transformation conserves the full-state purity.

For a finite projected dark-state ramp, Eq.~\eqref{eq:ramp} and the initial Gibbs state can both be expanded around the free canonical state $\rho_0=f(A)$. For a free-mode occupation $n_i$ the leading dynamical correction is
\begin{equation}
 \delta\langle n_i\rangle_{\rm ramp}^{(1)}
 =-\frac{i}{\hbar}\int_0^\tau\mathrm dt\,\operatorname{Tr}\bigl(\rho_0[n_i,V_I(t)]\bigr)=0,
 \label{eq:Sdyninvariant}
\end{equation}
where $V_I$ is the interaction-picture quartic perturbation. The free canonical state commutes with every free-orbital number operator, so the trace of this first-order commutator vanishes. The initial interacting Gibbs state already has a nonzero equilibrium occupation correction, evaluated in Appendix~\ref{sec:SPO}. Off-diagonal first-order corrections to $\gamma$ change the nondegenerate leading eigenvalue only at second order. Equation~\eqref{eq:retention} follows without treating the first-order input occupation shift as zero. The argument need not hold uniformly at exact leading-eigenvalue degeneracy.

The finite-strength calculation follows the same staged EIT trajectory, including its optical-photonic fraction $h(t)$, the quartic envelope $w(t)$, and total duration $1.25\,\mu$s. The initial \emph{thermal} spacing $x$ and the physical frequency $\Omega_D/(2\pi)$ are independent quantities: a small-$N$ temperature is adjusted to match the initial free occupation, not the physical oscillator frequency. At $r=0.4$, $T=10\,\mu\mathrm K$ gives $x\simeq0.146$ and $\Omega_D/(2\pi)\simeq0.0304$ MHz; the separate matched small-$N$ window uses $x\simeq0.888$ at about $1.64\,\mu\mathrm K$ with the \emph{same} physical frequency. The finite-$N=3$ simulation evaluates the complete interacting 56-dimensional Gibbs state and its unitary propagation.

For the matched-condensate case at $(r,\lambda_0)=(0.4,0.30)$, conversion changes the occupation fraction by about four parts per million. Refining the propagation grid from 160 to 320 steps changes the result by less than one part per billion. At weak reference coupling the conversion correction has leading quadratic behavior, consistent with Eq.~\eqref{eq:retention}. The occupation changes reported here arise from transverse dynamics within the projected dark manifold.

\begin{figure}[t]
\centering
\includegraphics[width=\columnwidth]{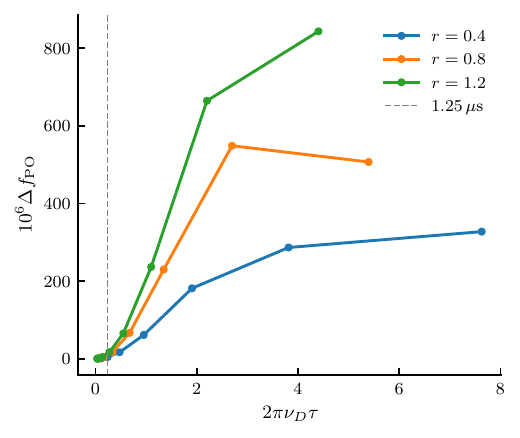}
\caption{Conversion duration and condensate retention in the $N=3$ projected dark-state model. Scaling the same EIT interaction-removal profile to different durations alters the conversion-induced change of $f_{\rm PO}$, shown in units of $10^{-6}$. The dashed line marks the reference $1.25\,\mu$s conversion duration in scaled time. Longer duration allows the transverse population to reorganize more strongly as the interaction is removed.}
\label{fig:Sduration}
\end{figure}

\subsection{Pair correlations beyond the one-body state}
\label{sec:Sm9null}
For the interacting $N$-boson state, one may construct a maximum-entropy quadratic fixed-number Gibbs control of the form
\begin{equation}
 \rho_{\rm ME}=\frac{\Pi_N\exp(-\sum_{ij}h_{ij}a_i^\dagger a_j)\Pi_N}{Z_N^{\rm ME}},
 \qquad \gamma_{ij}^{\rm ME}=\gamma_{ij}^{\rm int}.
 \label{eq:Smatchedgamma}
\end{equation}
The complete $6\times6$ one-body matrix, including off-diagonal elements, is matched numerically in the exact $N=3$ model. Reflection invariance gives no even--odd $x$ coherences in $\gamma$. Accordingly the quadratic generator can be block diagonal in single-orbital $x$ parity and conserves the number $N_{\rm odd,x}$ of bosons in odd-$x$ orbitals. The operator $a_{00}^\dagger a_{20}^\dagger a_{10}^2$ changes that number by two, so its expectation is exactly zero in the quadratic Gibbs reference. The interacting pair-scattering Hamiltonian conserves parity but need not conserve $N_{\rm odd,x}$, and hence may generate a nonzero pair expectation.

For the $r=0.4$, $T=10\,\mu\mathrm K$, $N=3$, $j\le2$ microscopic model at $\lambda=0.30$, a \emph{new quadratic control matched to the photon-output} one-body matrix has normalized output two-particle trace distance about $0.005$ from the interacting output. Matching the control to the photon-output one-body matrix accounts for any one-body reorganization during the conversion. The resulting two-body contrast characterizes this specified reference family. The pair comparison concerns correlations beyond a quadratic Gibbs reference; it is distinct from entanglement or phase-space nonclassicality. Passive mode mixing and mode-resolved coincidences offer candidate measurements of the postconversion pair coherence.

\section{Particle-number dependence}
\label{sec:SNscaling}

For a given $N$ the excited-state saturation scale is defined by Eq.~\eqref{eq:cross}. Using the representative normalization $N_*=200$ and $x_*=x_c^{(N_*)}\simeq0.097$, with fixed temperature and initial photonic fraction $f_\gamma=0.95$, the mass corresponding to that scale is $r_c(N)=[x_*f_\gamma/x_c^{(N)}]^2$. This scale locates the smooth finite-size crossover of the ideal-gas reference. The exact canonical ground-mode occupation fraction follows from the recurrence in Appendix~\ref{sec:Slinear}. The resulting crossover scales are listed in Table~\ref{tab:Ncrossover}.
\begin{table}[t]
\caption{Ideal-gas excited-state saturation reference for different fixed particle numbers. The crossover mass $r_c(N)$ uses the common reference mass normalization.}
\label{tab:Ncrossover}
\centering
\begin{tabular}{rcc}
\toprule
$N$ & $x_c^{(N)}$ & $r_c(N)$ \\
\midrule
20 & 0.32128 & 0.0823\\
40 & 0.22428 & 0.1688\\
80 & 0.15629 & 0.3476\\
120 & 0.12652 & 0.5304\\
200 & 0.09699 & 0.9025\\
300 & 0.07859 & 1.3748\\
400 & 0.06771 & 1.8519\\
\bottomrule
\end{tabular}
\end{table}

The physical comparison holds the bare interaction $g_m$ and Gaussian range $\sigma$ fixed. From Eq.~\eqref{eq:lambda}, $\lambda_N/(N-1)=f_m^2g_mJ_{0000}/(\hbar\Omega_D)$ is independent of $N$ at common cavity parameters. Thus $\lambda_N=(N-1)G(r,f_m)\lambda_{N_*}/(N_*-1)$, with $G$ from Eq.~\eqref{eq:SGorig}. Combining this identity with the free-canonical covariance formula yields Eq.~\eqref{eq:Nresponse}. Both the free occupations and their interaction-energy covariance depend on $N$, producing distinct size dependences for $\partial_{\lambda_N}f_{\rm PO}$ and the fixed-bare-coupling susceptibility $D_N$.

The $(N,r)$ comparison in Fig.~\ref{fig:Nscaling} spans $N=20$--$400$ and $0.4\le r\le1.2$ at $f_m=0.05$, keeping the microscopic Gaussian tensor and temperature fixed. All main results use oscillator shells through $j\le120$. Selected $j\le120$ versus $160$ comparisons give a maximum sampled relative discrepancy of approximately $0.18\%$ for the first-order susceptibility and $0.10\%$ for the ideal-gas condensate fraction. The quoted numerical agreement applies to these sampled parameter values and cutoffs.

A second controlled comparison adjusts the thermal parameter $x$ at each $N$ to obtain the same \emph{free} canonical condensate fraction (about $54.2\%$), while keeping the physical oscillator frequency and microscopic interaction fixed. At $r=0.4$, the fixed-bare-coupling derivatives $D_N$ for $N=(40,120,200,400)$ are approximately $(-0.0145,-0.0330,-0.0472,-0.0755)$ in this matched-condensate comparison. By contrast at fixed physical temperature they are approximately $(-0.00027,-0.0261,-0.0472,-0.0587)$. These values characterize the linear-response regime at the specified populations.

Under the ideal number-preserving mode map, $\gamma_{\gamma}=U^*\gamma_DU^{\mathsf T}$ proves invariance of all natural occupations at every fixed $N$. In the projected finite-time Hamiltonian $H_D(t)=\hbar\Omega_D(0)[h(t)A+\lambda_N w(t)S/(N-1)]$, the leading occupation correction is unchanged to first order in $\lambda_N$ for each fixed $N$. The coefficient of the higher-order correction can grow with particle number. In particular, for fixed microscopic $g_m$, $\lambda_N$ grows as $N-1$, so the domain of validity of the weak-interaction approximation changes with $N$.

In six full two-dimensional oscillator orbitals ($n_x+n_y\le2$), exact fixed-number projected conversion was performed for $N=2,3,4,5$ in Hilbert spaces of dimensions $21,56,126,252$. The thermal parameter was separately retuned to the same free condensate fraction at each $N$, while the \emph{physical} transverse frequency and original $1.25\,\mu$s EIT waveform were held fixed. The corresponding occupation changes are compared in Table~\ref{tab:Nconversion}.
\begin{table}[t]
\caption{Additional condensate-fraction change during the projected conversion in six-orbital few-body models. The two columns hold either the collective interaction parameter or the microscopic coupling fixed.}
\label{tab:Nconversion}
\centering
\begin{tabular}{rcc}
\toprule
$N$ & $\Delta f_{\rm PO}$ ($\lambda_N=0.30$) & $\Delta f_{\rm PO}$ (fixed $g_m$) \\
\midrule
2 & $7.74\times10^{-6}$ & $1.98\times10^{-10}$ \\
3 & $4.24\times10^{-6}$ & $4.39\times10^{-10}$ \\
4 & $3.09\times10^{-6}$ & $7.23\times10^{-10}$ \\
5 & $2.52\times10^{-6}$ & $1.05\times10^{-9}$ \\
\bottomrule
\end{tabular}
\end{table}
Here the fixed-$g_m$ column uses a reference $\lambda_{200}=0.30$ so $\lambda_N=0.30(N-1)/199$. In the fixed-$\lambda_N$ column, $g_m$ changes with particle number and these cases are not fixed-microscopic-coupling comparisons. Conversion was checked with 120 and 240 time steps; the largest absolute change in $\Delta f_{\rm PO}$ was $2.5\times10^{-9}$, and the fixed-bare-coupling changes had discretization errors below $4\times10^{-13}$. Exactness refers to the finite orbital Fock spaces under the projected dark-state Hamiltonian.

A separate notion of robustness is loss. With a mode-uniform passive channel of transmission $\eta$, the mean one-body density matrix changes as $\gamma\mapsto\eta\gamma$. Its normalized largest eigenvalue $\lambda_{\max}(\gamma)/\operatorname{Tr}\gamma$ is unchanged, but the all-particle no-loss probability for an initially fixed $N$ is $\eta^N$. The two diagnostics therefore quantify different aspects of survival. For mode-selective losses, even the normalized natural-orbital fraction need not remain invariant.

\section{Matter-assisted redistribution and intracavity isolation}
\label{sec:SRaman}
A number-conserving intermode process can be constructed from an auxiliary far-detuned optical state $|u\rangle$ and an external reservoir mode $b_\ell$. With an auxiliary detuning $\Delta_u$ large compared with its optical couplings and linewidth, elimination of the virtual excitation gives the effective vertex
\begin{equation}
 H_R=\hbar\sum_{q q'\ell}\Lambda_{qq'}^\ell
 a_q^\dagger S_{q'}b_\ell^\dagger+\mathrm{H.c.},
 \qquad \Lambda_{qq'}^\ell\sim \frac{g_q^{(u)}\xi_\ell}{\Delta_u}\mathcal I_{qq'},
 \label{eq:SRamanH}
\end{equation}
where $\mathcal I_{qq'}$ is the Raman spatial overlap. This is a schematic eliminated vertex, not a derivation of the spectral density and the interaction-dependent transition graph for a specific atom. The inverse dark--bright rotation gives
\begin{align}
 a_q^\dagger S_{q'}={}&-cs\,D_q^\dagger D_{q'}+c^2D_q^\dagger B_{q'}
 -s^2B_q^\dagger D_{q'}+sc B_q^\dagger B_{q'},
 \label{eq:SRamandecomp}
\end{align}
with $c^2=f_\gamma$, $s^2=f_m$ and $cs=\sqrt{f_\gamma f_m}$. The useful projected jump $D_q^\dagger D_{q'}$ conserves the total number $N_D$ but changes its transverse energy. Resolving the dark channel from the bright branches requires a characteristic redistribution rate $\Gamma_{\rm th}$ and reservoir bandwidth $B_R$ satisfying $\Gamma_{\rm th}\ll B_R\ll\Delta_{DB}/\hbar$, as well as sufficiently weak dispersive dark--bright mixing.

\begin{figure}[t]
\centering
\includegraphics[width=\columnwidth]{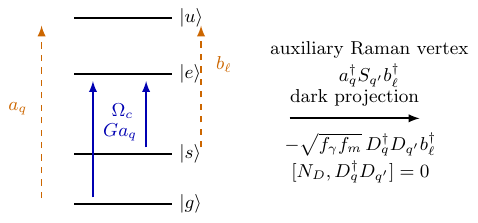}
\caption{Proposed auxiliary Raman redistribution channel. Cavity-photon and optical-control couplings form the EIT dark mode using $|g\rangle$, $|s\rangle$ and $|e\rangle$. A far-detuned auxiliary state $|u\rangle$ mediates a reservoir-assisted exchange that projects onto $D_q^\dagger D_{q'}$ within the dark manifold. The candidate vertex supplies free-DSP mode redistribution; interaction-resolved equilibration is described separately.}
\label{fig:SRaman}
\end{figure}

The auxiliary-level arrangement for this candidate exchange channel is illustrated in Fig.~\ref{fig:SRaman}.

For a harmonic transverse spectrum, odd Raman profiles proportional to $x$ or $y$ select neighboring oscillator levels with matrix elements proportional to $\sqrt{n_\alpha+1}$. A number-diagonal kinetic closure based on independent boson-enhanced transitions then has currents
\begin{align}
 F_{\nu+\hat\alpha\to\nu}&=\Gamma_\downarrow(n_\alpha+1)
 n_{\nu+\hat\alpha}(1+n_\nu),\\
 F_{\nu\to\nu+\hat\alpha}&=\Gamma_\uparrow(n_\alpha+1)
 n_\nu(1+n_{\nu+\hat\alpha}),
 \qquad \frac{\Gamma_\uparrow}{\Gamma_\downarrow}
 =e^{-\beta\hbar\Omega_D}.
 \label{eq:SRamancurrents}
\end{align}
The currents describe boson-enhanced redistribution between neighboring free-polariton modes. Their ratios refer to the \emph{free} dark-polariton ladder. The interacting Gibbs state of Eq.~\eqref{eq:intH} requires an interaction-resolved many-body generator. Interacting equilibration would require transition rates at the many-body energy differences and a graph that connects the relevant fixed-$N$ sectors; the abstract detailed-balance construction in Appendix~\ref{sec:Spreparation} describes that separate preparation assumption. In particular, the common Raman coupling can conserve center-of-mass/relative-motion sector structure; interactions modify the true many-body transition energies; and the kinetic Bose-profile distance compares one-body occupations with a reference distribution.

To compare intermode redistribution with depletion, the kinetic benchmark starts with 200 polaritons in transverse mode $(8,0)$ and includes uniform single-polariton loss $\kappa_{\rm DSP}$. The distance diagnostic was $D_{\rm BE}=N(t)^{-1}\sum_\nu|n_\nu(t)-n_\nu^{\rm BE}(t)|$, where the comparison profile uses the instantaneous surviving number and the dark-polariton spacing. In its retained $j_{\max}=80$ graph, a first-passage condition $D_{\rm BE}<0.05$ combined with mean population survival above $90\%$ required a redistribution-to-loss ratio of order $6\times10^2$ in the kinetic calculation, with $700$ used as a design illustration. At the redistribution threshold $D_{\rm BE}=0.04$, the mode-population closure gives the values in Table~\ref{tab:raman-kinetics}.
\begin{table}[t]
\caption{Mode-population redistribution and mean survival in the phenomenological Raman kinetic closure at $D_{\mathrm{BE}}=0.04$. These values describe free-polariton orbital occupations.}
\label{tab:raman-kinetics}
\centering
\begin{tabular}{ccc}
\toprule
Mass $r$ & Survival & $n_{00}/N(t)$\\
\midrule
1.2 & 0.919 & 0.032\\
1.0 & 0.915 & 0.060\\
0.8 & 0.909 & 0.141\\
0.6 & 0.926 & 0.323\\
0.4 & 0.955 & 0.553\\
\bottomrule
\end{tabular}
\end{table}
The condensed-side population grows as redistribution accelerates relative to loss. These mode occupations describe the free-spectrum kinetic closure; the interacting equilibrium state is defined independently by the energy-resolved many-body model. The many-body sections instead start from the explicitly conditional state in Eq.~\eqref{eq:rhoinput} and the compatible energy-resolved generator in Eq.~\eqref{eq:SL}.

\subsection{Coherent removal of atomic coupling}
\label{sec:Sisolation}
The intended sequence after preparing the DSP gas is
\begin{equation*}
\begin{gathered}
\text{reservoir off}\ \longrightarrow\ \Omega_c\text{ increase}\\
\longrightarrow\ \text{axial coupling reduction}\\
\longrightarrow\ G_{\rm res}\to0
\longrightarrow\Omega_c\to0.
\end{gathered}
\end{equation*}
In the illustrative schedule, reservoir switch-off takes $0.10\,\mu\mathrm{s}$, the $\Omega_c$ increase $0.20\,\mu\mathrm{s}$, ideal axial translation $1.00\,\mu\mathrm{s}$, and the terminal $G_{\rm res}$ ramp $0.05\,\mu\mathrm{s}$. The entire post-freeze sequence thus lasts $1.35\,\mu\mathrm{s}$, whereas the time-dependent projected Hamiltonian studied in Eqs.~\eqref{eq:ramp} and~\eqref{eq:kernel} follows only the final three active stages totaling $1.25\,\mu\mathrm{s}$. The optical and spatial stages define the prescribed control trajectory.

A finite-width cloud centered at a node still leaves a nonzero collective atom--cavity coupling. If $\Omega_c$ were switched off with $G_{\rm res}\ne0$, the instantaneous angle $\tan\theta=G_{\rm res}/\Omega_c$ would rotate the nominal dark branch back toward spin coherence rather than produce an isolated photon. The ideal endpoint therefore requires a final coherent reduction $G_{\rm res}(t)\to0$ while $\Omega_c$ remains finite. An example envelope is
\begin{equation}
\begin{aligned}
G_{\rm res}(t)&=\frac{G_{\rm res}(0)}{2}
\left[1+\cos\!\left(\frac{\pi t}{\tau_{\rm off}}\right)\right],\\
0&\le t\le\tau_{\rm off},\qquad
\tau_{\rm off}=0.05\,\mu\mathrm{s}.
\end{aligned}\label{eq:SGoff}
\end{equation}
At $G_{\rm res}=0$, the instantaneous dark operator is $D_q=a_q$. The subsequent $\Omega_c\to0$ operation then acts only on the decoupled atomic subsystem, leaving the intracavity photons as autonomous degrees of freedom. The final controlled reduction of $G_{\rm res}$ completes the decoupling, leaving an intracavity photon state at the endpoint.

When the spin-wave dispersion is nearly flat, the three-component single-particle EIT spectrum is almost homothetic to that of the photon oscillator: $E_{D,q}-E_{D,0}\simeq f_\gamma(E_{\gamma,q}-E_{\gamma,0})$. The three-component optical calculation gives relative deviations below $5\times10^{-6}$ through the thermally retained $j\le80$ manifold in its reference parameters. The deviation characterizes that single-particle spectrum over the retained orbital range. For ideal free-boson occupations that remain mode diagonal under the map, the same numbers can be interpreted as a free photon Bose profile with $T_{\rm out}\simeq T_{\rm in}/f_\gamma$ and appropriately transformed chemical potential. The temperature rescaling expresses a reversible change of the free-particle energy scale. The transfer of an interacting Gibbs state is generally a \emph{nonequilibrium} state relative to the final free photon Hamiltonian; its natural occupations are nevertheless preserved by the ideal mode map.

\subsection{Spatial mode preservation and photon loss}
\label{sec:Sspatial}
For an axial atomic density translated without changing its transverse profile, $n_a(\bm r,t)=n_\perp(\bm r_\perp)n_z[z-z_0(t)]$, the cavity-overlap matrix factorizes as $\mathcal M_{qq'}(t)=A_z(t)\mathcal M^\perp_{qq'}$. This gives $[\mathcal M(t),\mathcal M(t')]=0$. Mode-diagonal transport further requires $[\mathcal M^\perp,h_\gamma]=0$ or a quantitative bound on the off-diagonal couplings that connect distinct photon energies.

For a Gaussian axial atomic wavepacket with rms width $\sigma_z$ at center $z_0$, the overlap with the standing-wave intensity is
\begin{equation}
 A_z=\langle\cos^2(kz)\rangle=\frac12\left[1+e^{-2k^2\sigma_z^2}\cos(2kz_0)\right].
 \label{eq:SAz}
\end{equation}
At $z_0=\pi/(2k)+\delta z$, the sign of the cosine term reverses, leaving a finite node overlap. In a separable mode model the residual collective coupling ratio is
\begin{equation}
 \frac{G_{\rm node}}{G_{\rm anti}}=
 \left[\frac{1-e^{-2k^2\sigma_z^2}\cos(2k\delta z)}{1+e^{-2k^2\sigma_z^2}}\right]^{1/2}.
 \label{eq:SGnode}
\end{equation}
For the original ideal example with wavelength $556$ nm, displacement $\lambda_{\rm opt}/4\simeq139$ nm, and axial trap frequency $\omega_z/(2\pi)=500$ kHz, a harmonic packet with initial center at zero can be moved without terminal coherent excitation in half a trap period: shift the trap minimum initially to $d/2$, evolve under $z_c(t)=d[1-\cos(\omega_z t)]/2$ for $\tau_{\rm tr}=\pi/\omega_z=1.00\,\mu\mathrm{s}$, then place the minimum at $d$. This harmonic trajectory transports the wavepacket center without residual coherent oscillation in the ideal trap. For an assumed $^{171}$Yb axial packet at $10\,\mu$K, the model gives $\sigma_z\simeq8.4$ nm and a residual matter fraction near $1.2\times10^{-4}$ after node translation with doubled optical control, before the final coherent coupling removal. Its value is conditional on the model and its axial parameters.

A transverse displacement $\delta r_\perp$ induces oscillator-mode mixing. With $d_\perp=\delta r_\perp/a_{\rm ho}$ and $|n\rangle$ an oscillator state, the exact displaced-state survival probability is
\begin{equation}
 p_{n\to n}=e^{-d_\perp^2/2}\bigl[L_n(d_\perp^2/2)\bigr]^2,
 \qquad p_{\rm mix}=1-p_{n\to n}.
 \label{eq:Sjitter}
\end{equation}
Here $L_n$ is a Laguerre polynomial. The harmonic-oscillator calculation at $p_{\rm mix}=10^{-3}$ gives $d_\perp\simeq(1.09,0.699,0.497,0.353)\times10^{-2}$ at $n=(8,20,40,80)$. The values characterize individual high-shell orbitals; an interacting many-body preparation requires weighting the relevant occupied modes and full spatial coupling matrix.

Under an independent, mode-uniform one-particle attenuation model, the mean survival after the full $1.35\,\mu\mathrm{s}$ post-freeze schedule with $\kappa/(2\pi)=1$ kHz is $e^{-\kappa\tau}\simeq0.9916$, consistent with the original conditional accounting. The full notional mean survival includes the earlier kinetic stage and can be estimated by multiplying this factor by its closure-level freeze survival. The full fixed-$N$ no-loss probability is instead $e^{-N\kappa\tau}$, distinguishing survival of the many-body state from mean population retention. Mode-dependent losses can alter the natural-orbital spectrum itself. The three-component optical calculation also supplies a single-particle bright-state leakage benchmark for the chosen EIT trajectory. Together, axial overlap, transverse mode mixing, and photon loss determine how closely the optical evolution approaches the ideal isolated-state map.

\bibliographystyle{apsrev4-2}
\bibliography{reference}

@article{Greiner2002,
  author = {Greiner, M. and Mandel, O. and H\"ansch, T. W. and Bloch, I.},
  journal = {Nature},
  volume = {419},
  pages = {51},
  year = {2002},
  title = {Collapse and revival of the matter wave field of a Bose--Einstein condensate},
  doi = {10.1038/nature00968},
}

@article{Polkovnikov2011,
  author = {Polkovnikov, A. and Sengupta, K. and Silva, A. and Vengalattore, M.},
  journal = {Rev. Mod. Phys.},
  volume = {83},
  pages = {863},
  year = {2011},
  title = {Colloquium: Nonequilibrium dynamics of closed interacting quantum systems},
  doi = {10.1103/RevModPhys.83.863},
}

@article{Schemmer2018,
  author = {Schemmer, M. and Johnson, A. and Bouchoule, I.},
  journal = {Phys. Rev. A},
  volume = {98},
  pages = {043604},
  year = {2018},
  title = {Monitoring squeezed collective modes of a one-dimensional Bose gas after an interaction quench using density-ripple analysis},
  doi = {10.1103/PhysRevA.98.043604},
}

@article{Bougas2019,
  author = {Bougas, G. and Mistakidis, S. I. and Schmelcher, P.},
  journal = {Phys. Rev. A},
  volume = {100},
  pages = {053602},
  year = {2019},
  title = {Analytical treatment of the interaction quench dynamics of two bosons in a two-dimensional harmonic trap},
  doi = {10.1103/PhysRevA.100.053602},
}

@article{Carusotto2013,
  author = {Carusotto, I. and Ciuti, C.},
  journal = {Rev. Mod. Phys.},
  volume = {85},
  pages = {299},
  year = {2013},
  title = {Quantum fluids of light},
  doi = {10.1103/RevModPhys.85.299},
}

@article{BlochCarusottoWouters2022,
  author = {Bloch, J. and Carusotto, I. and Wouters, M.},
  journal = {Nat. Rev. Phys.},
  volume = {4},
  pages = {470},
  year = {2022},
  title = {Non-equilibrium Bose--Einstein condensation in photonic systems},
  doi = {10.1038/s42254-022-00464-0},
}

@article{Klaers2010NP,
  author = {Klaers, J. and Vewinger, F. and Weitz, M.},
  journal = {Nat. Phys.},
  volume = {6},
  pages = {512},
  year = {2010},
  title = {Thermalization of a two-dimensional photonic gas in a `white wall' photon box},
  doi = {10.1038/nphys1680},
}

@article{Klaers2010Nature,
  author = {Klaers, J. and Schmitt, J. and Vewinger, F. and Weitz, M.},
  journal = {Nature},
  volume = {468},
  pages = {545},
  year = {2010},
  title = {Bose--Einstein condensation of photons in an optical microcavity},
  doi = {10.1038/nature09567},
}

@article{Wang2019,
  author = {Wang, C.-H. and Gullans, M. J. and Porto, J. V. and Phillips, W. D. and Taylor, J. M.},
  journal = {Phys. Rev. A},
  volume = {99},
  pages = {031801(R)},
  year = {2019},
  title = {Theory of Bose condensation of light via laser cooling of atoms},
  doi = {10.1103/PhysRevA.99.031801},
}

@article{Pieczarka2024,
  author = {Pieczarka, M. and G\k{e}bski, M. and Piasecka, A. N. and Lott, J. A. and Pelster, A. and Wasiak, M. and Czyszanowski, T.},
  journal = {Nat. Photonics},
  volume = {18},
  pages = {1090},
  year = {2024},
  title = {Bose--Einstein condensation of photons in a vertical-cavity surface-emitting laser},
  doi = {10.1038/s41566-024-01478-z},
}

@article{Klaers2012PRL,
  author = {Klaers, J. and Schmitt, J. and Damm, T. and Vewinger, F. and Weitz, M.},
  journal = {Phys. Rev. Lett.},
  volume = {108},
  pages = {160403},
  year = {2012},
  title = {Statistical physics of Bose--Einstein-condensed light in a dye microcavity},
  doi = {10.1103/PhysRevLett.108.160403},
}

@article{Schmitt2014PRL,
  author = {Schmitt, J. and Damm, T. and Dung, D. and Vewinger, F. and Klaers, J. and Weitz, M.},
  journal = {Phys. Rev. Lett.},
  volume = {112},
  pages = {030401},
  year = {2014},
  title = {Observation of grand-canonical number statistics in a photon Bose--Einstein condensate},
  doi = {10.1103/PhysRevLett.112.030401},
}

@article{Kirton2013,
  author = {Kirton, P. and Keeling, J.},
  journal = {Phys. Rev. Lett.},
  volume = {111},
  pages = {100404},
  year = {2013},
  title = {Nonequilibrium model of photon condensation},
  doi = {10.1103/PhysRevLett.111.100404},
}

@article{Schmitt2015PRA,
  author = {Schmitt, J. and Damm, T. and Dung, D. and Vewinger, F. and Klaers, J. and Weitz, M.},
  journal = {Phys. Rev. A},
  volume = {92},
  pages = {011602(R)},
  year = {2015},
  title = {Thermalization kinetics of light: From laser dynamics to equilibrium condensation of photons},
  doi = {10.1103/PhysRevA.92.011602},
}

@article{Kirton2015,
  author = {Kirton, P. and Keeling, J.},
  journal = {Phys. Rev. A},
  volume = {91},
  pages = {033826},
  year = {2015},
  title = {Thermalization and breakdown of thermalization in photon condensates},
  doi = {10.1103/PhysRevA.91.033826},
}

@article{Schmitt2016PRL,
  author = {Schmitt, J. and Damm, T. and Dung, D. and Wahl, C. and Vewinger, F. and Klaers, J. and Weitz, M.},
  journal = {Phys. Rev. Lett.},
  volume = {116},
  pages = {033604},
  year = {2016},
  title = {Spontaneous symmetry breaking and phase coherence of a photon Bose--Einstein condensate coupled to a reservoir},
  doi = {10.1103/PhysRevLett.116.033604},
}

@article{Tang2024,
  author = {Tang, Y. and Dhar, H. S. and Oulton, R. F. and Nyman, R. A. and Mintert, F.},
  journal = {Phys. Rev. Lett.},
  volume = {132},
  pages = {173601},
  year = {2024},
  title = {Breakdown of temporal coherence in photon condensates},
  doi = {10.1103/PhysRevLett.132.173601},
}

@article{Fleischhauer2000,
  author = {Fleischhauer, M. and Lukin, M. D.},
  journal = {Phys. Rev. Lett.},
  volume = {84},
  pages = {5094},
  year = {2000},
  title = {Dark-state polaritons in electromagnetically induced transparency},
  doi = {10.1103/PhysRevLett.84.5094},
}

@article{Fleischhauer2002,
  author = {Fleischhauer, M. and Lukin, M. D.},
  journal = {Phys. Rev. A},
  volume = {65},
  pages = {022314},
  year = {2002},
  title = {Quantum memory for photons: Dark-state polaritons},
  doi = {10.1103/PhysRevA.65.022314},
}

@article{Fleischhauer2005RMP,
  author = {Fleischhauer, M. and Imamoglu, A. and Marangos, J. P.},
  journal = {Rev. Mod. Phys.},
  volume = {77},
  pages = {633},
  year = {2005},
  title = {Electromagnetically induced transparency: Optics in coherent media},
  doi = {10.1103/RevModPhys.77.633},
}

@article{Zimmer2008,
  author = {Zimmer, F. E. and Otterbach, J. and Unanyan, R. G. and Shore, B. W. and Fleischhauer, M.},
  journal = {Phys. Rev. A},
  volume = {77},
  pages = {063823},
  year = {2008},
  title = {Dark-state polaritons for multicomponent and stationary light fields},
  doi = {10.1103/PhysRevA.77.063823},
}

@article{Fleischhauer2008,
  author = {Fleischhauer, M. and Otterbach, J. and Unanyan, R. G.},
  journal = {Phys. Rev. Lett.},
  volume = {101},
  pages = {163601},
  year = {2008},
  title = {Bose--Einstein condensation of stationary-light polaritons},
  doi = {10.1103/PhysRevLett.101.163601},
}

@article{Kiffner2010,
  author = {Kiffner, M. and Hartmann, M. J.},
  journal = {Phys. Rev. A},
  volume = {82},
  pages = {033813},
  year = {2010},
  title = {Master equation approach for interacting slow- and stationary-light polaritons},
  doi = {10.1103/PhysRevA.82.033813},
}

@article{KimKim2026,
  author = {Kim, U.-S. and Kim, Y.-H.},
  journal = {npj Quantum Inf.},
  volume = {12},
  pages = {145},
  year = {2026},
  title = {Lieb--Liniger interaction via self-interacting stationary light polaritons},
  doi = {10.1038/s41534-026-01287-y},
}

@article{Kim2021,
  author = {Kim, B. and Chen, K.-T. and Hsiao, S.-S. and Wang, S.-Y. and Li, K.-B. and Ruseckas, J. and Juzeli\={u}nas, G. and Kirova, T. and Auzinsh, M. and Chen, Y.-C. and Chen, Y.-F. and Yu, I. A.},
  journal = {Commun. Phys.},
  volume = {4},
  pages = {101},
  year = {2021},
  title = {A weakly-interacting many-body system of Rydberg polaritons based on electromagnetically induced transparency},
  doi = {10.1038/s42005-021-00604-5},
}

@article{KimYu2023,
  author = {Kim, B. and Chen, K.-T. and Chen, K.-Y. and Chiu, Y.-S. and Hsu, C.-Y. and Chen, Y.-H. and Yu, I. A.},
  journal = {Phys. Rev. Lett.},
  volume = {131},
  pages = {133001},
  year = {2023},
  title = {Experimental demonstration of stationary dark-state polaritons dressed by dipole-dipole interaction},
  doi = {10.1103/PhysRevLett.131.133001},
}

@article{Chen2026,
  author = {Chen, D.-A. and Huang, K.-Y. and Hsu, C.-Y. and Xie, M.-C. and Yu, I. A. and Liao, W.-T.},
  journal = {Phys. Rev. Research},
  volume = {8},
  pages = {L032006},
  year = {2026},
  title = {Inhomogeneous mass trap for dark-state polaritons in atomic media},
  doi = {10.1103/g83q-29c9},
}

@article{PenroseOnsager1956,
  author = {Penrose, O. and Onsager, L.},
  journal = {Phys. Rev.},
  volume = {104},
  pages = {576},
  year = {1956},
  title = {Bose--Einstein condensation and liquid helium},
  doi = {10.1103/PhysRev.104.576},
}

@article{Balazs1998,
  author = {Balazs, N. L. and Bergeman, T.},
  journal = {Phys. Rev. A},
  volume = {58},
  pages = {2359},
  year = {1998},
  title = {Statistical mechanics of ideal Bose atoms in a harmonic trap},
  doi = {10.1103/PhysRevA.58.2359},
}

@article{WeissWilkens1997,
  author = {Weiss, C. and Wilkens, M.},
  journal = {Opt. Express},
  volume = {1},
  pages = {272},
  year = {1997},
  title = {Particle number counting statistics in ideal Bose gases},
  doi = {10.1364/OE.1.000272},
}

@article{Peyronel2012,
  author = {Peyronel, T. and Firstenberg, O. and Liang, Q.-Y. and Hofferberth, S. and Gorshkov, A. V. and Pohl, T. and Lukin, M. D. and Vuleti\'{c}, V.},
  journal = {Nature},
  volume = {488},
  pages = {57},
  year = {2012},
  title = {Quantum nonlinear optics with single photons enabled by strongly interacting atoms},
  doi = {10.1038/nature11361},
}

@article{Firstenberg2013,
  author = {Firstenberg, O. and Peyronel, T. and Liang, Q.-Y. and Gorshkov, A. V. and Lukin, M. D. and Vuleti\'{c}, V.},
  journal = {Nature},
  volume = {502},
  pages = {71},
  year = {2013},
  title = {Attractive photons in a quantum nonlinear medium},
  doi = {10.1038/nature12512},
}

@article{Doganov2013,
  author = {Doganov, R. A. and Klaiman, S. and Alon, O. E. and Streltsov, A. I. and Cederbaum, L. S.},
  journal = {Phys. Rev. A},
  volume = {87},
  pages = {033631},
  year = {2013},
  title = {Two trapped particles interacting by a finite-range two-body potential in two spatial dimensions},
  doi = {10.1103/PhysRevA.87.033631},
}

@article{Kubo1957,
  author = {Kubo, R.},
  journal = {J. Phys. Soc. Jpn.},
  volume = {12},
  pages = {570},
  year = {1957},
  title = {Statistical-mechanical theory of irreversible processes. I. General theory and simple applications to magnetic and conduction problems},
  doi = {10.1143/JPSJ.12.570},
}

@article{Glauber1963,
  author = {Glauber, R. J.},
  journal = {Phys. Rev.},
  volume = {130},
  pages = {2529},
  year = {1963},
  title = {The quantum theory of optical coherence},
  doi = {10.1103/PhysRev.130.2529},
}

\end{document}